\documentstyle[11pt,aaspp4]{article}
\def \cmsq           {\hbox{cm$^{-2}$}}
\def \deg          {\ifmmode ^{\circ}\else $^\circ$\fi}  
\def \etal         {{\it et~al.} }
\def \Hb          {\hbox{H$\beta$}}
\def \kms          {\rm{\hbox{km s$^{-1}$}}}
\def \lam          {$\lambda$}
\def \Lya          {\hbox{Ly$\alpha$}}
\def \Lyb          {\hbox{Ly$\beta$}}
\def \mum          {\hbox{$\mu$m}}
\def \ni           {\noindent}
\def \pcc           {\hbox{cm$^{-3}$}}

\def \zaz          {{$z_a\kern -1.5pt \approx\kern -1.5pt z_e$}}
\def \zgz          {{$z_a>z_e$}}
\def \zlz          {{$z_a<z_e$}}
\def \zllz         {{$z_a\kern -3pt \ll\kern -3pt z_e$}}
\def \Zsun         {\rm{\hbox{Z$_{\odot}$}}}           
\begin{document}
\pagestyle{plain}
\renewcommand{\baselinestretch}{2}
\thispagestyle{empty}

\ni{\Large\bf ELEMENTAL ABUNDANCES IN QSOS:}
\vspace{6pt}

\ni{\Large\bf Star Formation and Galactic Nuclear Evolution}

\ni{\Large\bf at High Redshifts}

\renewcommand{\baselinestretch}{1}
\bigskip\bigskip\bigskip
\ni{\large\it Fred Hamann}

\ni Department of Astronomy, University of Florida, 211 Bryant Space 
Sciences Center, Gainesville, FL 32611-2055; \ hamann@astro.ufl.edu (current)

\ni and

\ni Center for Astrophysics and Space Sciences, University of California, 
San Diego, La Jolla, CA 92093-0424
\bigskip\bigskip

\ni {\large\it Gary Ferland}

\ni Department of Physics and Astronomy, University of Kentucky, 
Lexington, KY 40506-0055; \ gary@pa.uky.edu 

\ni and

\ni Canadian Institute for Theoretical Astrophysics, University of 
Toronto, Toronto, ON, Canada, M5S 3H8
\bigskip\bigskip\bigskip

\smallskip

\ni Shortened Title: ELEMENTAL ABUNDANCES IN QSOS
\medskip\hrule
\newpage
\baselineskip 14pt

\begin{abstract}
Quasar (or QSO) elemental abundances provide 
unique probes of high-redshift star formation 
and galaxy evolution. There is growing evidence from both 
the emission and intrinsic absorption lines that QSO 
environments have roughly solar or higher metallicities 
out to redshifts $>$4. The range is not well 
known, but solar to a few times solar appears to be typical. 
There is also evidence for higher metallicities in 
more luminous objects, and for generally enhanced 
N/C and Fe/$\alpha$ abundances compared to solar ratios. 
\smallskip

These results identify QSOs with vigorous, 
high-redshift star formation -- consistent with the early 
evolution of massive galactic nuclei or dense proto-galactic 
clumps. However, the QSOs offer new constraints. For example, 1) 
most of the enrichment and star formation must occur before 
the QSOs ``turn on'' or become observable, on time scales 
of $\la$1~Gyr at least at the highest redshifts. 
2) The tentative result for enhanced Fe/$\alpha$ 
suggests that the first local star formation began at least 
$\sim$1~Gyr prior to the QSO epoch. 3) The star formation must 
ultimately be extensive in order to reach high metallicities, 
i.e. a substantial fraction of the local gas must 
be converted into stars and stellar remnants. The exact fraction 
depends on the shape of the initial mass function 
(IMF). 4) The highest derived metallicities require IMFs 
that are weighted slightly more toward massive stars 
than the in solar neighborhood. 5) High metallicities 
also require deep gravitational potentials. By analogy with 
the well-known mass--metallicity relation among low-redshift 
galaxies, metal-rich QSOs should reside in galaxies 
(or proto-galaxies) that are minimally as massive (or as 
tightly bound) as our own Milky Way.

\end{abstract}

\newpage
\tableofcontents
\newpage

\section{Introduction}

Quasi-stellar objects (QSOs or quasars) are valuable probes 
of the high-redshift Universe (Schneider 1998). Their 
most distant representatives are now measurable out to 
redshifts of $z\sim 5$ (Schneider, Schmidt \& Gunn 1991,  
Sloan Digital Sky Survey press release 1998). 
In Big Bang cosmologies, these redshifts correspond to times 
when the Universe itself was just $\sim$1~Gyr old (see Fig. 1). 

\begin{figure}[h]
\plotfiddle{figure1.ps}{3.2in}{0.0}{55.0}{55.0}{-225.0}{-478.0}
\end{figure}
\begin{quotation}
\noindent Fig. 1 --- Redshift versus age of the Universe 
in Big Bang cosmologies. The three solid curves correspond to 
$H_o = 65$~km~s$^{-1}$~Mpc$^{-1}$ and $\Omega_{\Lambda} = 0$ with 
$\Omega_{M} = 0$, 0.3 or 1. The dotted curve corresponds 
to $\Omega_{\Lambda} = 0.7$ and $\Omega_{M} = 0.3$. 
The ``error'' bars show the range of ages possible for $H_o$ between 
50 and 80~km~s$^{-1}$~Mpc$^{-1}$ (see Carroll \& Press 1992). 
\end{quotation}

Understanding the elemental abundances in these distant, 
early-epoch environments is a major goal of quasar research. 
Some of the first spectroscopic studies noted simply that quasar 
environments contain the usual array of ``metals'' 
(elements C, N, O and heavier) produced by stellar 
nucleosynthesis (Shklovskii 1965, Burbidge \& Burbidge 1967). 
More quantitative estimates of the abundances 
came later from theoretical work on the broad 
emission lines, culminating in the important  
review by Davidson \& Netzer (1979 --- hereafter DN79, 
also Baldwin \& Netzer 1978, Shields 1976). Those studies inferred 
solar or slightly higher metal abundances, with large uncertainties. 
The past two decades have seen considerable progress.   
Today we have a better theoretical understanding of quasar 
environments, and greater abilities to both observe and 
model a range of abundance diagnostics. 

We also have renewed motivation from the growing 
evidence that links quasars to galaxies. See, for example, Kormendy 
\etal (1998), Magorrian \etal (1998) for black hole--host galaxy 
mass correlations, Chatzichristou, Vanderriest \& Jaffe (1999), 
Hines \etal (1999), McLeod, Rieke \& Storrie-Lombardi (1999), 
Boyce, Disney \& Bleaken (1999), McLure \etal (1998), 
Aretxaga, Terlevich \& Boyle (1998), Carballo \etal (1998), 
Bahcall \etal (1997), Miller, Tran \& Sheinis (1996), 
McLeod \& Rieke (1995) for direct observations 
of QSO hosts, Cavaliere \& Vittorini (1998), 
Shaver \etal (1998), Terlevich \& Boyle (1993), 
Boyle \& Terlevich (1998), Osmer (1998)
for arguments based on QSO number-density evolution, 
McCarthy (1993), Saikia \& Kulkarni (1998), Haas \etal (1998), 
Brotherton \etal (1998a) for radio galaxy--radio quasar unification 
schemes, and Turner (1991), Haehnelt \& Rees (1993), 
Loeb \& Rasio (1994), Katz \etal (1994), Haehnelt, Natarajan 
\& Rees (1998), Haiman \& Loeb (1998), Taniguchi, Ikeuchi 
\& Shioya (1999) for theoretical 
links between QSOs and galaxy evolution.  
If quasars reside, as expected, in galactic nuclei 
or dense proto-galactic clumps, their abundances could yield 
unique constraints on the evolution of those environments. 
For example, quasar abundances can indirectly probe the 
star formation that came before QSOs, possibly the first 
stars forming in massive collapsed structures after the Big Bang. 
Other studies of high-redshift galaxies and metal enrichment, 
involving, for example, the ``Lyman-break'' objects (Steidel 
\etal 1998, Connolly \etal 1997) or the   
damped-\Lya\ or \Lya\ ``forest'' absorbers 
in QSO spectra (Pettini \etal 1997, Lu, Sargent \& Barlow 1998, 
Rauch 1998), probe more extended structures. 
The quasar results should therefore provide an important 
piece to the overall puzzle of high-redshift star formation 
and galaxy evolution. 

Here we review the status and implications of quasar abundance 
work. We regret that many interesting related topics must be 
excluded; in particular, we will consider the quasars themselves 
to be simply light sources surrounded by emitting and absorbing 
gas. We discuss three abundance diagnostics that are 
readily observable in QSOs at all redshifts: the broad emission 
lines (BELs), the broad absorption lines (BALs) and the 
intrinsic narrow absorption lines (NALs). We include  
just these ``intrinsic'' spectral features to probe the 
abundances near QSO engines --- 
excluding measures of the extended host galaxies, nearby 
cluster galaxies or cosmologically intervening gas. 
We begin with separate discussions of each abundance probe (\S\S2--3), 
followed by a summary of the overall results (\S4).  
We then consider the plausible enrichment schemes, making a case 
for normal chemical evolution by stars in galactic 
nuclei (\S5). Within that scheme, we use results from galactic 
studies (\S6) to derive further implications of the QSO abundances 
(\S7). We close with a brief outline for future work (\S8). 

In several sections below we will present results of 
photoionization calculations performed with the 
numerical code Cloudy (version 90.05, Ferland \etal 1998). 
This code is freely available on the world wide web 
(http://www.pa.uky.edu/$\sim$gary/cloudy/). Finally, we will define 
solar abundances according to the meteoritic results in 
Grevesse \& Anders (1989).

\newpage
\section{Emission Line Diagnostics}

\subsection{Overview}

Quasars are surprisingly alike in their emission-line spectra 
(Osmer \& Shields 1999 and refs. therein); 
for example, the range of intensity ratios is far less than 
in galactic nebulae. Figure 2 
shows a composite UV spectrum that is fairly typical of QSOs 
without strong BALs. The object-to-object similarities span the 
full range of QSO redshifts, $0.1\la z\la 5$, more than 4 orders 
of magnitude in luminosity, and billions of years in cosmological 
look-back time. The emission lines are either insensitive 
to the metal abundances, or QSOs have similar abundances across  
enormous ranges in other parameters. We will argue that 
the truth involves a bit of both explanations. 

We will focus on the BELs in the rest-frame UV because they are 
present and relatively easy to measure in all QSOs at all redshifts. 
Furthermore, unlike the narrow emission lines, there is no ambiguity 
about their close physical connection to QSO engines (DN79).

\begin{figure}[h]
\plotfiddle{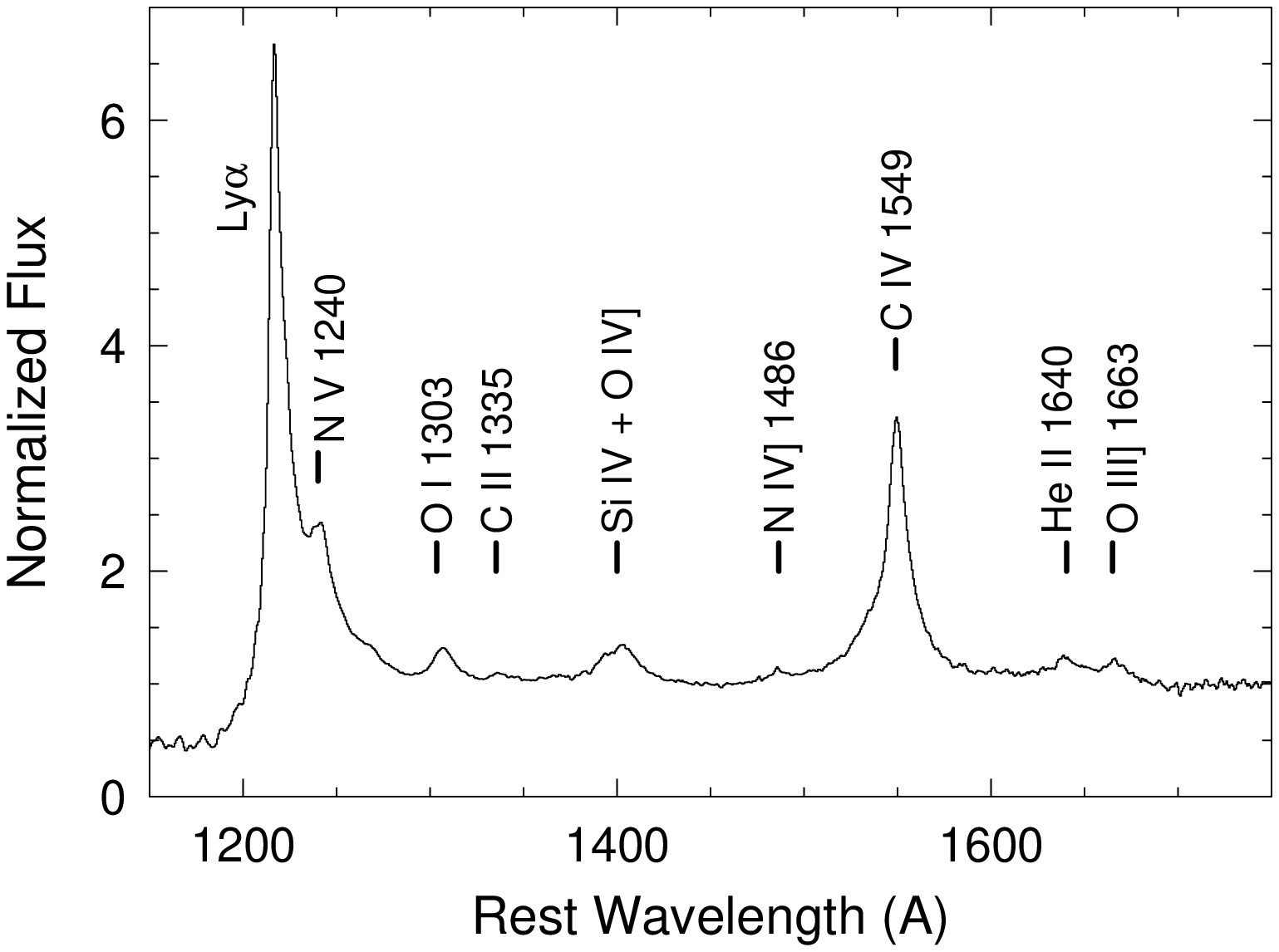}{2.9in}{0.0}{75.0}{75.0}{-235.0}{-198.0}
\end{figure}
\begin{quotation}
\ni Fig. 2 ---  Normalized mean spectrum of 13 QSOs at $z>4$ 
(from Shields \etal 1997). Prominent BELs are labeled. 
\end{quotation}

\subsection{Origin of the Broad Emission Lines}

Quasar emission-line research is an example of the ``inverse problem'' 
in astrophysics.  We know the answer --- the observed spectrum of a 
quasar, and we are trying to understand the question --- the conditions 
that created it. Any model of the line-forming regions will 
have uncertainties related to uniqueness, but these can be 
minimized by considering the astrophysical context and by 
limiting the models to essential properties. The essential 
properties of the BEL region (BELR) are as follows:

1) The BELR is photoionized. 
The main evidence for photoionization is that the 
emission-line spectra change in response to changes in
the continuum, with lag-times corresponding to characteristic 
radii of the BELR (Peterson 1993). The shape of the
ionizing continuum is a fundamental parameter and is in 
itself an area of active
research (e.g. Zheng \etal 1997; Korista, Baldwin \& Ferland 1997a, 
Brunner \etal 1997, Laor 1998).  
We will present calculations using simple power-laws between 
1~\mum\ and 100~keV, and describe results that do not depend 
strongly on the continuum shape.

2) The BELR spans a range of distances from the central
object. The line variability or ``reverberation'' studies just mentioned 
find different lag-times for different ions. Highly ionized species 
tend to lie closer to the continuum source. Overall, 
the radial distances scale with luminosity, such that 
$R\approx 0.1(L/10^{46} {\rm ergs\ s}^{-1})^{1/2}$~pc is a typical 
value (Peterson 1993). 

3) The BELR has a wide range of densities and ionization states. 
The range in ionization follows simply from the lines detected, 
from OI~\lam 1303 to at least NeVIII~\lam 774 (Hamann \etal 
1998). The range in density comes mainly from 
the estimated radii and photoionization theory (e.g. 
Ferland \etal 1992). Clouds\footnote{We use the term 
``cloud'' loosely, referring to some localized part of the 
BELR but not favoring any particular model or geometry (see Arav 
\etal 1998, Mathews \& Capriotti 1985).} with densities from $10^8$ 
to $>$$10^{12}$~\pcc\ may be present. Any given object could 
have a broad mixture of BELR properties (Baldwin \etal 1995, 1996).

4) The BELR probably has large column densities. 
Large columns, typically $N_H\ga 10^{23}$~\cmsq , 
were originally used in BELR simulations 
to produce a wide range of ionizations in single clouds 
(Kwan \& Krolick 1981 --- hereafter KK81, 
Ferland \& Persson 1989). These large columns might 
not apply globally because we now know that different lines 
form in different regions. In our calculations below, we will 
truncate the clouds at the hydrogen recombination front, 
with the result that different clouds/calculations can have 
different total column densities. However, the 
truncation depths are in all cases large enough to 
include the full emission regions of the relevant lines.

5) Thermal velocities within clouds are believed to dominate the 
local line broadening and radiative transfer. The observed 
line-widths are thus due entirely to bulk motions of the gas. 
This issue is important because {\it i)} 
continuum photoexcitation (``pumping'') can overwhelm other 
excitation processes if the local line broadening 
(e.g. micro-turbulence) is large, and {\it ii)} the line optical depths 
and thus photon escape probabilities (see below) vary inversely 
with the amount of line broadening. The interplay between these 
factors makes it hard to predict the behavior of a given line without 
explicit calculations. Shields, Ferland \& Peterson 
(1995) plot some examples for the particular case of low column 
density clouds. One argument against significant 
micro-turbulence involves the \Lya /\Hb\ intensity ratio. 
Simple recombination theory predicts a ratio of about 34 
(Osterbrock 1989 --- hereafter O89) 
while the observed value is far smaller, closer to 10 (Baldwin
1977a). This discrepancy is worsened by micro-turbulence 
(Ferland 1999). The solution probably requires severely 
trapped \Lya\ photons resulting from large optical depths at 
thermal line widths (see also Netzer \etal 1995). 

\subsection{Strategies for Abundance Work.}

There is much that is unknown about QSO line-forming regions. 
We do not, for example, have a clear picture of the overall 
geometry or the spatial variations of key parameters; 
but we do not need this information for abundance work. 
The emission lines from photoionized clouds are controlled 
fundamentally by the energy balance and microphysics. 
The strategy for abundance studies is to identify line 
ratios that have significant abundance sensitivities and minimal 
dependences on other unknown or uncertain parameters. 
For example, we can minimize the sensitivity to 
large-scale geometric effects by comparing lines that form 
as much as possible in the same gas. 
Detailed simulations are often needed to identify 
useful line ratios and quantify their parameter sensitivities. 
Simple analytic expressions can be used for some 
applications and they can help, in any case, provide physical 
insight into the emission-line behaviors. 

Below we review some of the basic principles of photoionization 
and emission-line formation. See O89 and Mihalas (1978) for 
further reviews,  Davison \& Netzer (1979 --- hereafter DN79), 
KK81, Ferland \& Shields 1985, and Netzer (1990 --- hereafter N90) 
for applications to QSOs, and Ferland \etal (1998) for 
more on the numerical simulations and input atomic data. 

\subsection{Basics of Abundance Analysis}

\subsubsection{Collisionally-Excited Lines}

Collisionally-excited lines form by the internal 
excitation of an ion following electron impact. Their emissivities, 
or energy released per unit volume and time, follow from the 
statistical equilibrium of the energy levels. For example, the 
equilibrium (detailed balance) equation for a 2-level atom is, 
\begin{equation}
n_l n_e q_{lu}\ =\ n_u(\beta A_{ul}+ n_e q_{ul}) \ \ \ \ 
[{\rm cm}^{-3}\ {\rm s}^{-1}]
\end{equation}
where $n_e$ is the electron density, 
$\beta$ is the probability for line photons 
escaping the local region ($0\leq\beta\leq1$), $A_{ul}$ is the 
spontaneous decay rate, $n_u$ and $n_l$ are the number densities 
in the upper and lower states, and $q_{lu}$ and 
$q_{ul}$ are the upward and downward collisional rate coefficients, 
respectively. Note that $\beta\sim\tau^{-1}$ when $\tau\gg 1$, where 
$\tau$ is the line-center optical depth (Frisch 1984). 
For most applications the ions are mainly in their ground state 
and $n_l$ is approximately the ionic density. The line 
emissivity is, 
\begin{equation}
\epsilon_{coll}\ =\ n_u\beta A_{ul} h\nu_o\ =\ n_l\beta A_{ul} h\nu_o
\left({{n_e q_{lu}}\over{\beta A_{ul}+n_e q_{ul}}}\right)
\ \ \ \ [{\rm ergs\ cm}^{-3}\ {\rm s}^{-1}]
\end{equation}
where $\nu_o$ is the line frequency. 
This emissivity has a strong temperature dependence because 
$q_{ul}\propto T^{-1/2}$ and $(q_{lu}/q_{ul}) =   
(g_u/g_l)\exp{(-h\nu_o/kT)}$, where $g_u$ and $g_l$ are the 
statistical weights. In the high 
density limit we have, 
\begin{equation}
\epsilon_{coll}\ =\  n_l\beta A_{ul} h\nu_o 
\,{{g_l}\over{g_u}}
\,\exp{\left(-{{h\nu_o}\over{kT}}\right)}
\end{equation}
and the levels are said to be ``thermalized.'' Line thermalization, 
where $\epsilon_{coll}$ no longer depends on the transition strength, 
additionally requires $\tau\gg 1$. ($A_{ul}$ and $\tau$ are 
both proportional to the oscillator strength, which 
therefore drops out of the factor $\beta A_{ul}\approx 
A_{ul}/\tau$ in Eqn. 3 if $\tau\gg 1$.) 
At low densities we have,
\begin{equation}
\epsilon_{coll}\ =\  n_l n_e q_{lu} h\nu_o\ \propto\ 
n_l n_e T^{-1/2}\,\exp{\left(-{{h\nu_o}\over{kT}}\right)}
\end{equation}
Note that $\epsilon_{coll}$ scales here like the density squared, 
compared to the linear dependence in Equation 3. 
The critical density, $n_{crit}$, between these two limits is the 
density where the two terms in the denominator of Equation 2 are equal, 
\begin{equation}
n_{crit}\ =\  {{\beta A_{ul}}\over{q_{ul}}}\approx 
{{A_{ul}}\over{\tau q_{ul}}}\
\end{equation}
where the approximate relation holds only if $\tau\gg 1$. 
Physically $n_{crit}$ is the density where the upper level is as likely 
to be de-excited by collisions as by radiative decays. Note that 
significant optical depths have the effect of 
lowering $n_{crit}$. Also note that 
transitions with very different oscillator strengths (but similar 
collision strengths) will have similar $n_{crit}$ in 
the limit $\tau\gg 1$ (because $A_{ul}/\tau$ is independent of 
oscillator strength). 

\subsubsection{Recombination Lines}

The most prominent recombination lines belong to HI, HeI and 
HeII, with HI \Lya\ being typically strongest. These lines form 
by the capture of free electrons into excited states, followed 
by radiative decays to lower states.  In
the simplest case, where every photon escapes freely and 
competing processes are unimportant, the emissivity is,
\begin{equation}
\epsilon_{rec}\ =\ n_{i}\,n_e\,\alpha_{rad}\,h\nu_o\ \propto\ 
n_{i}\,n_e T^{-1} \ \ \ \
[{\rm ergs\ cm}^{-3}\ {\rm s}^{-1}]
\end{equation}
where $\alpha_{rad}$ is the radiative recombination coefficient 
into the upper energy state and $n_{i}$ is the number density of 
parent ions. The temperature dependence is approximate and 
derives from $\alpha_{rad}$ (see O89). 

\subsubsection{Deriving Abundance Ratios}

These two types of lines can be combined to form three 
types of ratios for abundance analysis. 
The general idea is that for any element $a$ in ion stage $i$, 
the observed line 
intensity, $I(a_i)$, is proportional to the density in that 
ion, $n(a_i)$, times a function of the overall gas 
density and temperature, $F(a_i,T,n)$, such that 
$I(a_i) = n(a_i) F(a_i,T,n)$. The ionic abundance ratios 
are then given by,
\begin{equation}
{{n(a_i)}\over{n(b_j)}}\ =\ {{I(a_i)}\over{I(b_j)}}\, {{F(b_j,T,n)}\over{F(a_i,T,n)}}
\end{equation}
Abundance studies require line pairs for which the ratio of the two 
functions $F$ is nearly constant or has limiting behaviors 
that still allow for abundance constraints.
The last step is to convert the ionic abundances
into elemental abundances, which we express logarithmically 
relative to solar ratios as\footnote{Our notation here 
is based on the usual definition of logarithmic abundances 
normalized to solar ratios, $[a/b]\equiv \log(a/b) - \log(a/b)_{\odot}$.},
\begin{equation}
\left[{a\over b}\right]\ = \ 
\log\left({{n(a_i)}\over{n(b_j)}}\right)\ +\
\log\left({{f(b_j)}\over{f(a_i)}}\right)\ +\
\log\left({b\over a}\right)_{\odot} 
\end{equation}
where $f(a_i)$ is the fraction of element $a$ in ion stage $i$, 
etc. The middle term on the right hand side is the ionization 
correction ($IC$), which can be deduced from numerical simulations 
or set to zero (in the log) based on the similarity of the species 
(Peimbert 1967). Another strategy is to compare summed 
combinations of lines from different ion stages so that 
$IC$ tends to zero on average (Davidson 1977). 

Ratios of pure recombination lines are simplest 
because they are least sensitive to the temperature and 
density. In principle, we could derive the He/H abundance from 
these ratios. However, in practice, all of the strong 
HI and HeI recombination lines in QSOs, 
most notably \Lya , are affected by collisions 
and thermalization effects. Moreover, because H$^o$, 
He$^+$ and He$^{+2}$ have different ionization energies, 
they need not be co-spatial in the BELR and their levels 
of ionization depend on the different numbers of 
photons available to produce each ion (Williams 1971). 
As a result, the H and He recombination spectra are most 
useful as indicators of the shape of the ionizing continuum 
(e.g. Korista \etal 1997a). We do not expect substantial 
deviations from solar He/H abundances anyway, based on 
normal galactic chemical evolution, and the BEL 
data are grossly consistent with that expectation. 

The second possible ratio involves collisional to recombination
lines. These ratios have strong temperature dependences (compare 
Eqns. 3 and 4 to Eqn. 6). Nonetheless, they can still be used 
for abundance work if the temperature sensitivities are 
quantified by explicit calculations. For example, there is an upper 
limit on the line ratio NV~\lam 1240/HeII~\lam 1640 related to the 
maximum temperature attained in photoionized BELRs. 
That upper limit sets a firm lower limit on the N/He abundance 
(\S2.6.3 below).

The last ratio, and the one most often used, involves two 
collisionally-excited lines. Roughly a dozen collisionally-excited
BELs are routinely measured in the UV spectra of quasars, so there 
are a variety of possibilities. The ideal collisionally-excited line
pair would have similar excitation energies, so their
ratio has a small $h\Delta\nu_o /kT$ and thus a small temperature 
dependence (Eqns. 3 and 4). 
Similar values of $n_{crit}$ and similar ionization energies 
further minimize the sensitivities to density and BELR structure. 
Well-chosen ratios that meet these criteria can sometimes 
provide abundance estimates without recourse 
to detailed simulations (e.g. Shields 1976, \S2.6.1 below)

\subsection{Photoionization Simulations}

A photoionized cloud is essentially a large-scale fluorescence 
problem. Energy comes into the cloud via continuum radiation, 
is converted into kinetic energy by the photo-ejection of
electrons, and then leaves the cloud by various emission processes 
-- mainly line radiation. The lines are thus the primary coolants; 
their total intensity depends on energy conservation and not 
at all on particular cloud properties. 

In general situations, e.g. dense environments like 
BELRs, individual line strengths can be governed by a number 
of competing processes and by feedback related to the 
cloud structure and energy balance. Detailed 
calculations are needed to simultaneously consider a complex 
network of coupled processes. Here we describe some basic 
results for the line formation and ionization structure 
in realistic BELR clouds. 

\subsubsection{Parameters of Photoionization Equilibrium}

The fundamental parameters in photoionization simulations 
are the shape and intensity of the ionizing continuum, and 
the gas' space density, column density, and chemical composition. 
The flux of hydrogen-ionizing photons at the illuminated 
face of a cloud is, 
\begin{equation}
\Phi(H)\ \equiv\ \int_{\nu_{LL}}^{\infty}{{f_{\nu}}\over{h\nu}}
\,d\nu \ \ \ \ [{\rm photons\ cm}^{-2}\ {\rm s}^{-1}]
\end{equation}
where $f_{\nu}$ is the energy flux density and $\nu_{LL}$ is 
the frequency corresponding to 1 Ryd. A dimensionless ionization 
parameter $U\equiv\Phi(H)/cn_H$ is often used instead, where $c$ 
is the speed of light and $n_H$ is the total hydrogen density 
(H$^o$ + H$^+$). 
$U$ is proportional to the level of ionization and has the
advantage of stressing homology relations between clouds with 
the same $U$ but different $\Phi(H)$ and $n_H$. This simplification 
is appropriate if we are interested in just the gross 
ionization structure or in emission lines that are not 
collisionally suppressed. More generally, we can use either 
$\Phi(H)$ or $U$ as long as the density is also specified. 

\subsubsection{A Computed Structure}

Figure 3 shows the ionization structure of a typical BELR cloud 
photoionized by a power-law spectrum with $\alpha = -1.5$, 
where $f_{\nu}\propto \nu^{\alpha}$. 
The hydrogen recombination front occurs at a depth of 
$\sim$10$^{12}$~cm, while the He$^{+2}$--He$^+$ front is near
10$^{11}$~cm. Note that there is significant ionization 
beyond the nominal H$^o$--H$^+$ front, due to penetrating X-rays 
and Balmer continuum photoionizations out of the $n=2$ level 
in H$^o$ (KK81). Some important low-ionization lines like 
FeII form in that region. The ionization fractions in plots like 
Figure 3 help us identify ions, 
such as O$^{+5}$, N$^{+4}$ and He$^{+2}$, that are roughly 
co-spatial and thus good candidates for abundance comparisons. 

\begin{quotation}
\centerline{XXXXXX INSERT FIGURE HERE XXXXXX}
\noindent Fig. 3 --- Ionization structure for a 
nominal BELR cloud with $n_H = 10^{10}$~\pcc , $\log U = -1.5$ and 
solar abundances.
\end{quotation}

\subsubsection{An Example: the CIV~\lam 1549 Equivalent Width}

CIV~\lam 1549 is one of the strongest collisionally-excited lines 
in quasar spectra. The left panel of Figure 4 shows how its  
predicted equivalent width changes with the 
density ($n_H$) and ionizing flux ($\Phi(H)$, see Korista \etal 
1997b for many more similar plots). Powerful selection effects are 
clearly at work; the line radiates efficiently over just a 
narrow range of parameters.  Varying $\Phi(H)$ is equivalent to
moving the cloud closer or farther from the continuum source.  
The line is weak at large values of $\Phi(H)$, because 
carbon is too highly ionized, and at low values of $\Phi(H)$,
because carbon is too neutral. The line strength also changes with 
the gas density. When the density is above $n_{crit}$, the line is
collisionally suppressed and other permitted lines take over 
the cooling.  When the density is low, 
the line weakens as the many forbidden and semi-forbidden lines 
become efficient coolants and the gas temperature declines. The line 
is most prominent at $n_H\approx 10^{10}$~\pcc\ and $\log U\approx -1.5$, 
which are the canonical BELR parameters deduced over twenty
years ago from analysis of the CIV emission (DN79).

\begin{quotation}
\centerline{XXXXXX INSERT FIGURE HERE XXXXXX}
\noindent Fig. 4 --- Predicted equivalent width (EW) 
of CIV~\lam 1549 as a function 
of the cloud density, $n_H$, and incident ionizing 
flux, $\Phi(H)$. The equivalent 
width here is dimensionless (line flux/$\nu_o f_{\nu_o}$ in the continuum) 
and applies for the hypothetical case of global covering factor 
unity. Flux ratios for NV~\lam 1240/HeII~\lam 1640 and NV/CIV are 
also shown. Other parameters are the same as Fig. 3.
\end{quotation}

It is important to remember that these selection effects exist 
whenever we observe an emission line.  Baldwin \etal (1995)
showed that a typical quasar BEL spectrum might result simply 
from selection effects operating in BELRs with 
a wide range of cloud properties (e.g. density and distance 
from the QSO). Numerical simulations can identify pairs of 
lines with similar selection behaviors so that their ratios 
are insensitive to the ranges or specific values of the parameters.

\subsubsection{Line Dependence on Continuum Shape}

Figure 5 shows a series of calculations using different 
incident spectral shapes. 
The actual shape of the ionizing continua 
in QSOs is a complicated issue, but the UV to X-ray slopes 
are roughly consistent with $\alpha\sim -1.5$, near the center 
of the range shown (see Laor 1998, 
Korista \etal 1997a for recent discussions). The results 
in Figure 5 mainly reflect the conservation of energy in the cloud. 
Harder spectra (less negative $\alpha$) provide more heating per 
photoionization, leading to higher temperatures. The increased 
heating requires more line cooling via 
collisionally-excited lines like CIV. The ratio of a 
collisionally-excited line to a recombination line, 
such as CIV/\Lya , is proportional to the cooling per recombination 
or equivalently the heating per photoionization (DN79). 
Such ratios therefore have a strong continuum-shape dependence. 
The strengths of collisionally-excited lines relative to the 
adjacent continuum (i.e. their equivalent widths) also depend 
on the spectral slope because of the temperature sensitivity 
and because the continuum below the lines might be very 
different from that controlling the ionization. Ratios of 
collisionally-excited lines, such as NV/CIV, can similarly 
depend on the spectral shape if their ionization or 
excitation energies are different. In dense BELRs, these 
simple behaviors can be moderated by other effects. 
For example, the \Lya\ equivalent width increases 
with spectral hardening at fixed $U$ (Fig. 5) because it 
has a significant collisional (temperature-sensitive) 
contribution. 

\begin{quotation}
\centerline{XXXXXX INSERT FIGURE HERE XXXXXX}
\noindent Fig. 5 --- Predicted line flux ratios, gas 
temperatures ($T_4 = T/10^4$~K in the O$^{+2}$ zone, 
i.e. weighted by the O$^{+2}$ 
fraction), and dimensionless equivalent 
widths in \Lya\ (EW, as in Fig. 4) are plotted for clouds 
photoionized by different power law spectra. 
Other parameters are the same as Fig. 3. The lines 
are CIII~\lam 977, NIII~\lam 991, OVI~\lam 1034, 
NV~\lam 1240, CIV~\lam 1549, HeII~\lam 1640, 
OIII]~\lam 1664, NIII]~\lam 1750 and CIII]~\lam 1909. 
\end{quotation}

\subsubsection{Line Dependence on Abundances}

The left-hand panel of Figure 6 shows a series of calculations 
for clouds with different metallicities, $Z$ (scaled from solar 
and preserving solar ratios). The strengths of the 
collisionally-excited lines relative to \Lya\ change little 
with $Z$. In particular, CIV/\Lya\ varies negligibly for 
$0.1\la Z\la 30$~\Zsun\ (see also Hamann \& Ferland 1993a, 
hereafter HF93a). We have already noted that these ratios 
are more sensitive to the continuum shape (\S2.5.4). 
Their lack of sensitivity to $Z$ can be traced to feedback 
in the energy balance. As the metal abundances grow, 
the line cooling increases. The growing metallicities, 
which might otherwise increase the metal line strengths, are 
thus balanced in real clouds by lower temperatures --- 
with the result that the total metal line flux stays 
constant. This feedback is especially important for 
strong lines, like CIV, that by themselves control a large 
fraction of the cooling. Weak lines respond better 
to abundance changes. At low metallicities ($Z\la 0.02$~\Zsun ) 
none of the metal lines are important coolants and their overall 
strengths do scale with $Z$. 

Another factor in the line 
behaviors at high $Z$ is the increasing bound-free 
continuum absorption by metal ions.
The metals absorb a larger fraction of 
the far-UV flux at high $Z$, such that the H and He recombination 
lines become somewhat weaker. This effect 
dominates the high-$Z$ rise in OVI/HeII and NV/HeII in 
Figure 6. 

The right-hand panel in Figure 6 shows the same line 
ratios as before, but in this 
case nitrogen is scaled such that 
N/H~$\propto Z^2$ (where N/H is solar at $Z=$~\Zsun ). 
This selective scaling is based on the expected secondary 
nucleosynthesis of nitrogen (\S6 below). 
Shields (1976) noted that this abundance pattern should  
occur in QSOs by analogy with its direct observation 
in galactic HII regions. Figure 6 shows that 
it leads to a strong metallicity dependence for line ratios 
involving nitrogen. This strong dependence is possible 
because the N lines do not control the cooling. 

\newpage
\begin{quotation}
\centerline{XXXXXX INSERT FIGURE HERE XXXXXX}
\noindent Fig. 6 --- Predicted line flux ratios for 
photoionized clouds 
with different metallicities, $Z$. The metals are 
scaled together (preserving solar ratios) in the left-hand 
panel, while nitrogen is selectively scaled like 
$Z^2$ in the right panel. 
Other parameters are the same as Fig. 3. See Fig. 5 
for line notations.  
\end{quotation}

\subsection{Abundance Diagnostics and Results}

\subsubsection{Intercombination Lines}

Shields (1976) proposed using various collisionally-excited 
intercombination (semi-forbidden) lines to derive 
metal-to-metal abundance ratios in QSOs. He emphasized the strengths 
of NIII]~\lam 1750 and NIV]~\lam 1486 compared 
to OIII]~\lam 1664, CIII]~\lam 1909 and CIV~\lam 1549 
as potential diagnostics of the overall 
metallicity. As noted above, the metallicity dependence stems 
from the expected $Z^2$ scaling of N via secondary 
nucleosynthesis (also \S6 below). Shields selected 
lines with similar ionization and excitation energies, so 
that their ratios are insensitive to the uncertain 
temperature, ionization and geometry. 
Comparisons with the measured line ratios in QSOs 
(see also Davidson 1977, Baldwin \& Netzer 1978, DN79, 
Osmer 1980, Uomoto 1984) suggested that N/C and N/O are 
often solar or higher, consistent with solar or higher 
metallicities. Gaskell, Shields \& Wampler (1981) extended this 
analysis to SiIII]~\lam 1892 and other lines to show that the 
refractory elements cannot be substantially depleted by dust 
in BELRs. 

One drawback of the  
intercombination lines is that most of them 
are weak and therefore difficult (or impossible) to measure. 
Nonetheless, the best recent measurements (Wills \etal 
1995, Laor \etal 1995, Boyle 1990, Baldwin \etal 1996) 
support the earlier results. 
It is now possible to gather even more data for these lines 
at a range of redshifts. A note of caution is that the 
strong feature generally attributed to CIII]~\lam 1909 can have 
large contributions from other lines (Laor \etal 1995, 1997, 
Baldwin \etal 1996), so 
that ratios like NIII]/CIII] might systematically 
underestimate N/C if line blending is not accounted for. 

A more serious concern is that 
the early theoretical work did not consider the range 
of high densities now believed to be present in the 
BELR (\S2.2). The intercombination lines probably 
form at or near their critical densities (typically 
$3\times 10^9$ to 10$^{11}$~\pcc\ for $\beta = 1$ in Eqn. 5). 
Lines with different $n_{crit}$ could have different degrees of 
collisional suppression. (For example, the calculated results 
using $n_e\approx n_H = 10^{10}$~\pcc\ 
in Figs. 5 and 6 favor large NIII]/CIII] at a given N/C abundance 
because CIII] is collisionally suppressed above its 
$n_{crit}\approx 3\times 10^9$~\pcc .) If there is a range 
of densities, lines with different $n_{crit}$ might form 
in different regions (even if they have similar ionizations),  
leading to a geometry dependence. 
Nonetheless, line ratios involving similar $n_{crit}$ and 
similar sensitivities to other parameters, such as 
NIII]/OIII], could still be 
robust abundance indicators when they are measurable. 
More theoretical work is needed to explore the parameter 
sensitivities and selection effects that can influence  
these lines in complex BELRs.  

\subsubsection{Permitted Lines}

There are several possibilities for abundance diagnostics 
among the permitted UV lines. Figure 6 shows that 
NIII~\lam 991/CIII~\lam 977 
and NV~\lam 1240/CIV~\lam 1549 should be good tracers of N/C. 
Another possibility is NV/HeII~\lam 1640, or perhaps 
NV/(CIV+OVI~\lam 1034). The NV, OVI and HeII lines form in 
overlapping regions (Fig. 3), as do NIII and CIII, so their 
flux ratios should be insensitive to the global BELR structure. 
Also, as noted above, the N lines are not important coolants 
and thus responsive to abundance changes. 
There are practical problems with most of these lines, however; 
NV is blended with \Lya , CIII, NIII and HeII are weak, and CIII, NIII 
and OVI lie in the ``forest'' of intervening \Lya\ absorption lines. 
Nonetheless, improvements in the quality of data (for example, 
high resolution and high signal-to-noise spectra in the \Lya\ 
forest) are permitting increasingly accurate measurements of 
these lines in large QSO samples.

\subsubsection{NV/HeII and NV/CIV}

Some of the first studies of large QSO samples 
noted that NV~\lam 1240 is often stronger than 
predicted by photoionization 
models using solar abundances (Osmer \& Smith 1976 and 1977). 
The NV/HeII and NV/CIV ratios have since received particular
attention as abundance diagnostics (Hamann \& Ferland 1992, 
HF93a, Hamann \& Ferland 1993b --- hereafter HF93b, 
and Ferland \etal 1996 --- hereafter F96). 
Figure 7 shows the measured 
ratios in these lines for QSOs at different redshifts 
(from HF93b and Hamann \etal 1997a, with some new data and 
modifications based mainly on Wills \etal 1995 and Baldwin \etal 
1996). NV/HeII is the ratio of a collisional to recombination line, 
with the expected strong temperature dependence (\S2.4). 
Calculations similar to (but more exhaustive than) those shown in 
Figures 4--6, indicate that NV/HeII reaches a maximum value 
linked to the maximum temperature in photoionized clouds (F96). 
The maximum NV/HeII ratio is $\sim$2--3 for solar N/He abundances, 
depending on how ``hard'' a continuum shape one considers 
realistic for QSOs. (Beware that the highest ratios in Fig. 4 
occur for parameters where both lines are growing weak, 
cf. the EW(CIV) plot or Korista \etal 1997b.) 
Nominal BELR parameters predict NV/HeII near 
unity for solar N/He (Figs. 4--6). These 
predictions fall well below most of the measured ratios (Fig. 7), 
implying that QSOs typically have super-solar N/He. 
The ad hoc (high) temperatures that 
would be needed to explain the observed NV/HeII ratios with 
solar N/He are inconsistent with photoionization equilibrium, and 
would lead to strong far-UV emission lines. The fact that these 
far-UV line strengths are not seen sets an upper limit on the 
temperature and supports the result for super-solar N/He (F96).

The NV/CIV lines are collisionally excited with similar 
energies, so the temperature dependence 
is smaller than NV/HeII. Nominal BELR 
parameters predict NV/CIV of order 0.1 for solar N/C (Figs. 
4--6, also HF93a,b). Comparisons with the data in Figure 7 
thus indicate super-solar N/C for most QSOs. 

The two NV ratios together therefore imply that 
1) quasar metallicities are often solar or higher, 
especially in high redshift, high luminosity objects, 
and 2) nitrogen (e.g. N/C) is typically enhanced compared 
to solar ratios (F96, HF93a,b). 
The conclusion for enhanced N/C is based largely 
on NV/CIV, but we note that the scaling of $N\propto Z^2$ leads to 
self-consistent estimates of $Z$ based on NV/HeII and NV/CIV (Fig. 7). 
The actual $Z$ values are uncertain, but the main point is 
that many observed ratios require $Z\ga$~\Zsun . 
Figures 6 and 7 combined suggest that the nominal 
metallicity range is $1\la Z\la 10$~\Zsun\  
for standard photoionization parameters and $N\propto Z^2$. 

\begin{figure}[h]
\plotfiddle{figure7.ps}{4.0in}{0.0}{60.0}{60.0}{-246.0}{-516.0}
\end{figure}
\begin{quotation}
\noindent Fig. 7 --- Measured NV/HeII and NV/CIV flux 
ratios versus redshift (left panels) and 
continuum luminosity (right). The upper and lower ranges 
might be undersampled (especially for NV/HeII at redshifts $>$1) 
because limits on weak lines (e.g. HeII) were often 
not available from the literature. The two asterisks in each 
panel represent mean values measured by Osmer \etal (1994) 
for ``high'' and ``low'' luminosity QSOs at redshift $>$3. 
The solid curves are predictions 
based on chemical evolution models (discussed in \S6 below). 
\end{quotation}

HF93b noted that the observed NV ratios tend to be higher in 
more luminous sources (Fig. 7). 
Most BELs exhibit the well-known ``Baldwin effect," that is, 
lower equivalent widths at higher luminosities (Baldwin 1977b). 
This effect is well established in CIV and appears to be even 
stronger in OVI (Zheng \etal 1995, 
Kinney, Rivolo \& Koratkar 1990, Osmer \& Shields 1999). 
Surprisingly, NV does not show this effect (Osmer \etal 1994, 
Laor \etal 1995, Francis \& Koratkar 1995) 
even though its ionization is intermediate 
between CIV and OVI and its electron structure is identical. 
We proposed that the peculiar NV behavior is due 
to generally higher metallicities and more enhanced N abundances 
in more luminous QSOs. The recent theoretical study of the 
Baldwin effect by Korista, Baldwin \& Ferland (1998) gives
quantitative support to that conclusion. In \S7 we will 
argue that this proposed metallicity--luminosity trend in 
QSOs could naturally result from a mass--metallicity 
correlation among their host galaxies. 

The abundance results based on NV have been questioned 
by Turnshek \etal (1988, 1996) and Krolik \& Voit (1998), who 
argue that the NV BEL forms largely by resonance scattering in 
an outflowing BAL region. NV might be selectively enhanced 
by this mechanism because it can scatter both the continuum 
and the underlying \Lya\ emission line. However, explicit 
calculations of the line scattering (Hamann, Korista \& Morris 
1993, Hamann \& Korista 1996, Hamann, 
Korista \& Ferland 1999a) do not support this scenario. 
For example, 1) the amount of NV scattering estimated by 
Krolik \& Voit (1998) is too large by a factor of 
$\sim$3 on average, because BALRs do not 
generally have the right velocity/optical depth structure 
to scatter all of the incident \Lya\ photons. In particular, 
NV BALs are not usually black across the \Lya\ emission line. 
2) It is difficult for NV ions in high-velocity 
BAL winds to scatter \Lya\ photons into simple emission 
profiles with observed half-widths of typically 2000 to 2500~\kms . 
For example, isotropic scattering of the \Lya\ flux would produce 
BEL half-widths of $\sim$6000~\kms\ (the velocity separation 
between the NV and \Lya\ lines). Anisotropic scattering 
(e.g. in BALRs with equatorial or bipolar geometries) would lead 
to strong orientation effects and systematically broader BEL 
profiles in BAL versus non-BAL QSOs. These differences are 
not observed (Weymann \etal 1991). 3) It is not clear 
why, in individual spectra, the NV emission profiles 
should closely resemble those of other BELs if the former is 
produced by scattering in a high-velocity BAL wind while the 
latter are collisionally excited in a separate 
region (i.e. the usual BELR --- whose velocity field is not mostly 
radial based on the reverberation studies, T\"urler \& 
Courvoisier 1997, Korista \etal 1995). Finally, 4) 
large scattering contributions to NV would minimally require 
much larger global BALR covering factors (the fraction 
of the sky covered by the BALR as seen from the central QSO) 
than expected from their observed detection frequency 
in (randomly oriented) QSO samples. Goodrich (1997) and  
Krolik \& Voit (1998) argue that larger global covering 
factors could occur, but that issue is not settled.  

Another concern is that complex BELR geometries 
might cause the NV/HeII and NV/CIV abundance indicators 
to fail --- but they would fail in opposite directions. 
Specifically, 
clouds that are truncated at different physical depths (see Fig. 3) 
could produce strong HeII with little or no NV and CIV emission, 
or strong HeII and NV with little or no CIV. For a given 
abundance set, this type of truncation could therefore either 
lower the observed NV/HeII ratio or increase NV/CIV. 
Comparing the data to simulations that do not 
take truncation into account (Figs. 4--6) might then 
lead to underestimated N/He abundances or overestimated 
N/C. However, we have already shown that these two line ratios 
yield similar metallicities when compared to the non-truncated 
simulations, so we are not likely being 
mislead by complex BELR geometries. Moreover, the 
NV/HeII ratio provides in any case a secure lower limit on N/He.

\subsubsection{FeII/MgII}

The broad FeII emission lines pose unique problems  
because the atomic physics is complex and many 
blended lines contribute to the spectrum, 
particularly at the wavelengths $\sim$2000--3000~\AA\ and 
$\sim$4500--5500~\AA . Nonetheless, FeII is worth the effort  
because a delay of $\sim$1~Gyr in the Fe enrichment, 
relative to $\alpha$ elements such as O, Mg or Si,  
might provide a ``clock'' for constraining
the ages of QSOs and the epoch of their first star formation 
(see \S\S6--7 below, also HF93b).  

A series of important papers on FeII emission 
(Osterbrock 1977, Phillips 1977,1978, Grandi 1981) culminated 
with Wills \& Netzer (1983) and Wills, Netzer \& Wills (1985, 
hereafter WNW). They performed sophisticated calculations 
showing that the large observed FeII fluxes, 
e.g. FeII(UV)/MgII~\lam 2799, require that either Fe is 
several times overabundant (compared to solar ratios) 
or some unknown process dominates the FeII excitation. 
One process that might selectively enhance FeII emission 
is photoexcitation by \Lya\ photons (Johansson \& Jordan 1984). 
The absorption of \Lya\ radiation can pump electrons from 
the lower (metastable) energy levels of Fe$^+$ into specific 
high-energy states, leading to 
fluorescent cascades. WNW discounted 
this mechanism because it appeared insignificant in their 
simulations, but Penston (1987) noted that 
\Lya\ pumping is known to be important in some emission-line  
stars, such as the symbiotic star RR Tel, and therefore 
might be important in QSOs. More recent FeII simulations 
using better atomic data and exploring a wider range of 
physical conditions 
(Sigut \& Pradhan 1998, Verner \etal 1999) suggest that \Lya\ 
can be important in some circumstances, but it is not yet 
clear if those circumstances occur significantly in QSOs. 

Recent observations have renewed interest in 
this question by showing that the FeII(UV)/MgII emission 
fluxes can be larger than the WNW predictions even at 
$z>4$, with the tentative conclusion that Fe/Mg is at 
least solar (and thus the objects are at least $\sim$1~Gyr old,  
Taniguchi \etal 1997, Yoshii, Tsujimoto \& Kawara 
1998, Thompson, Hill \& Elston 1999 and refs. therein). 
New theoretical efforts, 
such as Sigut \& Pradhan (1998) and Verner \etal (1999), 
are needed to test these conclusions and quantify the 
uncertainties. However, a better way to measure 
Fe/$\alpha$ might be with the intrinsic NALs (see below). 

\section{Absorption Line Diagnostics}

\subsection{Overview: Types of Absorption Lines}

Quasar absorption lines can have a variety of intrinsic 
or cosmologically intervening origins.  
We exclude from our discussion the 
damped-\Lya\ absorbers and the ``forest'' of many narrow \Lya\ 
systems with weak or absent metal lines because 
they form in cosmologically intervening gas (Rauch 1998). 
The remaining metal-line systems can be divided into 
two classes according to  
their broad or narrow profiles. This division is 
a gross simplification, but still useful 
because it distinguishes the clearly intrinsic broad 
lines from the many others of uncertain origin. Here  
we briefly characterize the two (broad and narrow) line types. 

\subsubsection{Broad Absorption Lines (BALs)}

Broad absorption lines 
are blueshifted with respect to the emission lines and have 
velocity widths of at least a few thousand \kms\ 
(for example, Fig. 8). They appear in 10 to 15\% of 
optically-selected QSOs and clearly identify 
high-velocity winds from the central engines. 
The precise location of the absorbing gas is unknown, 
but there is little doubt that it is intrinsic --- originating 
within at least a few tens of parsecs from the QSOs.  
See recent work by Weymann \etal (1991), Barlow \etal (1992), 
Korista \etal (1993), Hamann \etal (1993), 
Voit, Weymann \& Korista (1993), Murray \etal (1995), 
Arav (1996), Turnshek \etal (1997), 
Brotherton \etal (1998b), and the reviews by 
Turnshek (1988, 1994), Weymann, Turnshek \& Christianen (1985) 
and Weymann (1994, 1997). 

\begin{figure}[h]
\plotfiddle{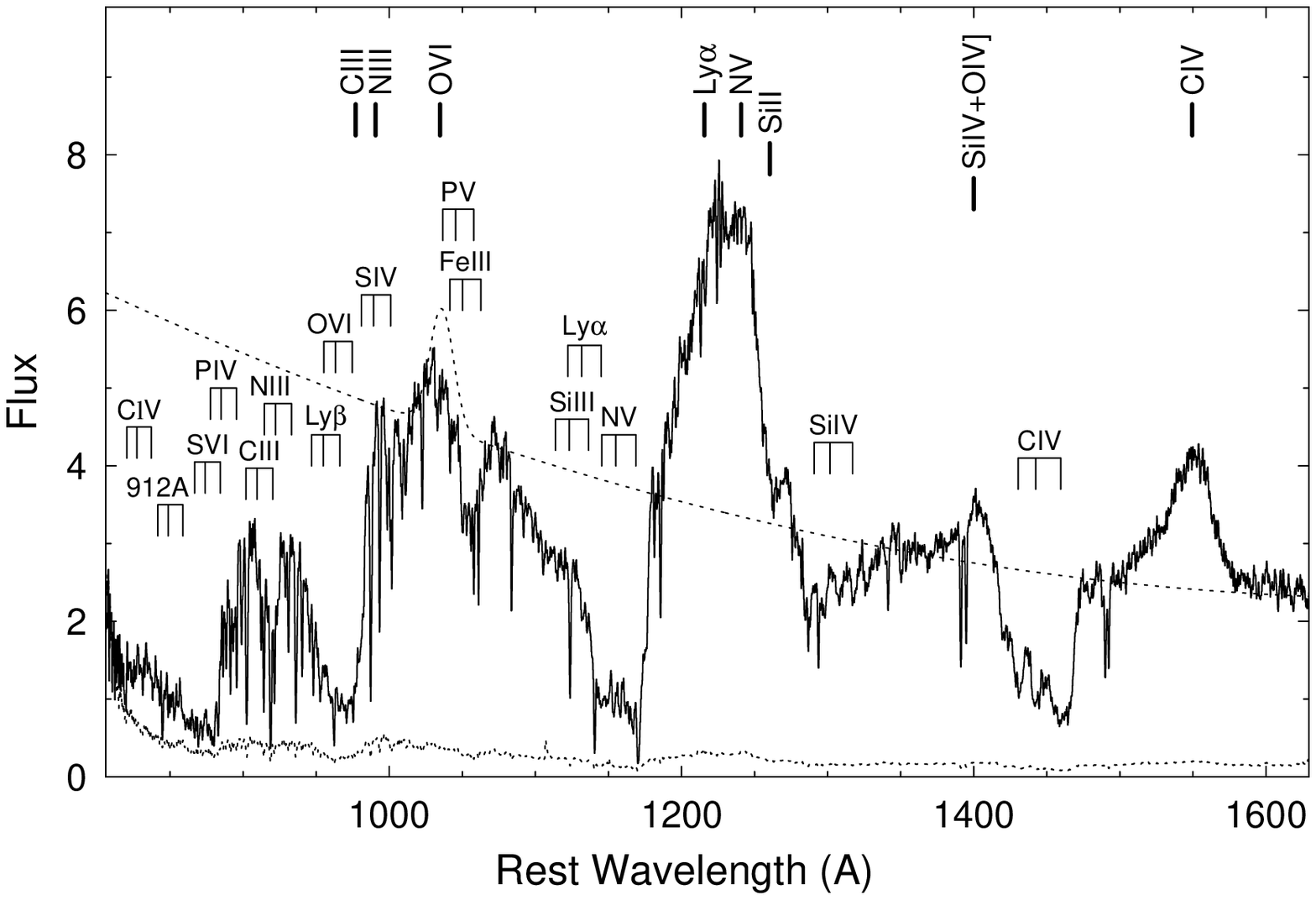}{2.5in}{0.0}{65.0}{65.0}{-205.0}{-200.0}
\end{figure}
\begin{quotation}
\noindent Fig. 8 --- Spectrum of the BALQSO PG~1254+047 
(emission redshift $z_e = 1.01$) with 
emission lines labeled across the top and possible BALs marked 
below at redshifts corresponding to the 3 deepest minima in the 
CIV trough. Not all of the labeled lines are detected. The smooth 
dotted curve is a power-law continuum fit extrapolated to short 
wavelengths (from Hamann 1998). 
\end{quotation}

\subsubsection{Narrow Absorption Lines (NALs)}

A practical definition of NALs would limit their full 
widths at half minimum (FWHMs) to 
less than the velocity separation of important 
doublets (e.g. $<$500~\kms\ for CIV, $<$1930 \kms\ for SiIV 
or $<$960 \kms\ for NV), because it is our ability to 
resolve these doublets that makes their analysis 
fundamentally different from the BALs (\S3.2.2 below). 

NALs can form in a variety of locations, ranging from 
very near QSOs, as in ejecta like the BALs, to unrelated gas 
or galaxies at cosmological distances (Weymann \etal 1979). 
It is not yet known what fraction of 
NALs at any velocity shift meet our definition of intrinsic 
(\S1). Several studies have noted a statistical excess 
of NALs within a few thousand \kms\ of the emission redshifts. 
These are the so-called ``associated'' or \zaz\ 
absorbers (with redshifts close to the emission 
redshift, Weymann \etal 1979 and 1981, Young \etal 1982, 
Foltz \etal 1986 and 1988, Anderson \etal 1987). 
Their strengths and frequency of occurrence 
appear to correlate with the QSO luminosities or radio 
properties, suggesting some physical relationship (also M\"oller 
\etal 1994, Aldcroft, Bechtold \& Elvis 1994, Wills \etal 1995, 
Barthel, Tytler \& Vestergaard 1997). 
These correlations may extend to NALs at blueshifts 
of 30,000~\kms\ or more (Richards \& York 1998). 
Nontheless, we might expect a larger 
fraction of intrinsic NALs nearer the emission redshift 
and, if they are ejected from QSOs, they should appear  
at \zlz\ rather than \zgz . 

Several tests have been developed 
to help identify intrinsic NALs, including 1) 
time-variable line strengths, 2) multiplet ratios that imply 
partial line-of-sight coverage of the background light 
source(s), 3) high gas densities inferred from excited-state 
absorption lines, and 4) well-resolved line profiles that are 
smooth and broad compared to both thermal line widths 
and to the velocity dispersions expected in intervening clouds 
(e.g. Bahcall, Sargent \& Schmidt 1967, Williams \etal 1975, 
Young \etal 1982, Barlow \& Sargent 1997, Hamann \etal 1997b 
-- hereafter H97b, Hamann \etal 1997c, Petitjean \& Srianand 1999, 
Ganguly \etal 1999, and refs. therein). 
These criteria might not be definitive individually, 
but they sometimes appear in combination. 

Figure 9 shows a \zaz\ NAL system that is clearly intrinsic 
based on time-variable line strengths, partial line-of-sight 
coverage and relatively broad profiles. 
High metallicities might be another indicator of intrinsic 
absorption (\S3.4 below), but that criterion would bias abundance 
studies; we would like to determine 
the intrinsic versus intervening nature independent 
of the abundances. The other (non-abundance) 
tests indicate that intrinsic NALs can have 
velocity shifts out to $\ga$24,000~\kms\ and a wide range of  
FWHMs down to $\la$30~\kms . See the references above and  
the reviews of \zaz\ systems 
by Weymann \etal (1981) and Foltz \etal (1988).  

\subsection{General Abundance Analysis}

Abundance estimates from absorption lines are, in principle, 
more straightforward than for emission lines because the 
absorption strengths are not sensitive to the temperatures or 
space densities. Moreover, absorption 
lines yield direct measures of the column densities in different 
ions. We need only apply appropriate ionization corrections 
to convert the column densities into relative abundances. 
For example, the abundance ratio of any two elements $a$ and 
$b$ can be written, 
\begin{equation}
\left[{a\over b}\right]\ = \ 
\log\left({{N(a_i)}\over{N(b_j)}}\right)\ +\
\log\left({{f(b_j)}\over{f(a_i)}}\right)\ +\
\log\left({b\over a}\right)_{\odot} 
\end{equation}
which is identical to Equation 8 except that the $N$ here are the 
column densities. Once again we define the ionization correction 
as $IC\equiv \log(f(b_j)/f(a_i))$. 
Abundance studies would ideally compare lines with similar 
ionizations to minimize $IC$ and reduce the sensitivity 
to potentially complex geometries. Unfortunately, the lines 
available often require significant ionization corrections. 
In particular, we are often forced to compare 
highly-ionized metals (such as CIV) to HI (\Lya ) to derive the 
metallicity. We must therefore use ionization models. 
 
\begin{figure}[h]
\plotfiddle{figure9a.ps}{2.7in}{0.0}{38.0}{38.0}{-163.0}{-285.0}
\plotfiddle{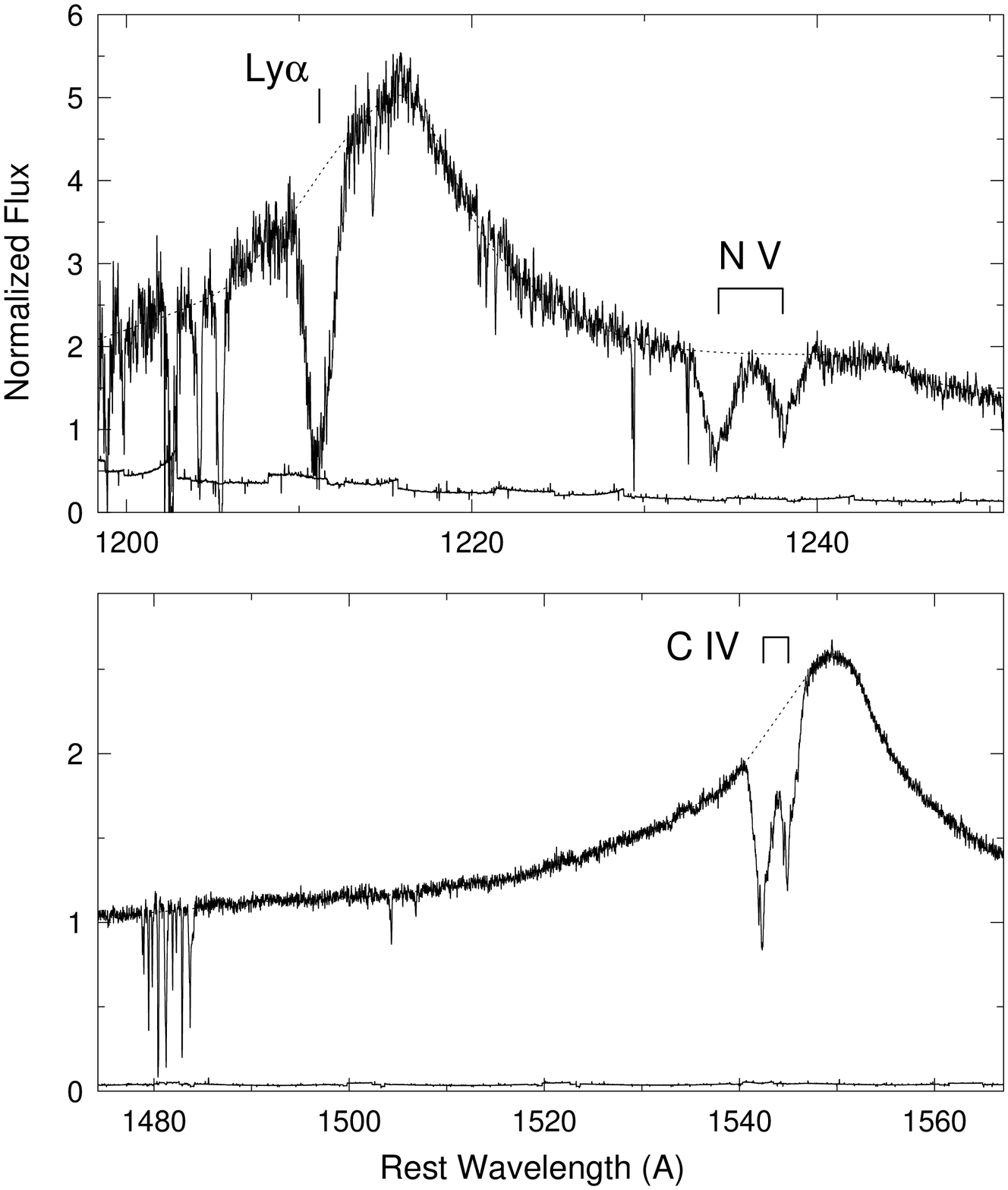}{2.7in}{0.0}{38.0}{38.0}{-115.0}{-30.0}
\end{figure}
\begin{quotation}
\noindent Fig. 9 --- Spectra of the \zaz\ absorber 
in UM675 ($z_e = 2.15$) showing its time-variability in two 
epochs (top panel) and broad, smooth profiles at higher 
spectral resolution (9~\kms , bottom panels, 
from Hamann \etal 1995, 1997b).
\end{quotation}

The usual assumption is that the gas is photoionized 
by the QSO continuum flux. Collisional ionization would 
lead to lower derived metallicities because it creates 
less HI (and more HII) for a given level of ionization 
in the metals (cf. Figs. 4 and 5 in Hamann \etal 1995, 
also Turnshek \etal 1996). However, collisional ionization 
has been generally dismissed for BAL regions (BALRs) 
because 1) it would 
be energetically hard to maintain, 2) it would 
produce excessive amounts of line emission (because 
of the much higher temperatures), and 3) it is hard to 
reconcile with the observed simultaneous variabilities in BAL 
troughs across a wide range of velocities (Weymann \etal 1985, 
Junkkarinen \etal 1987, Barlow 1993). In contrast, the 
strong radiative flux known to be present in QSOs 
provides a natural ionization source. We will assume 
that photoionization dominates in both BALRs and 
intrinsic NALRs. 

\begin{figure}[h]
\plotfiddle{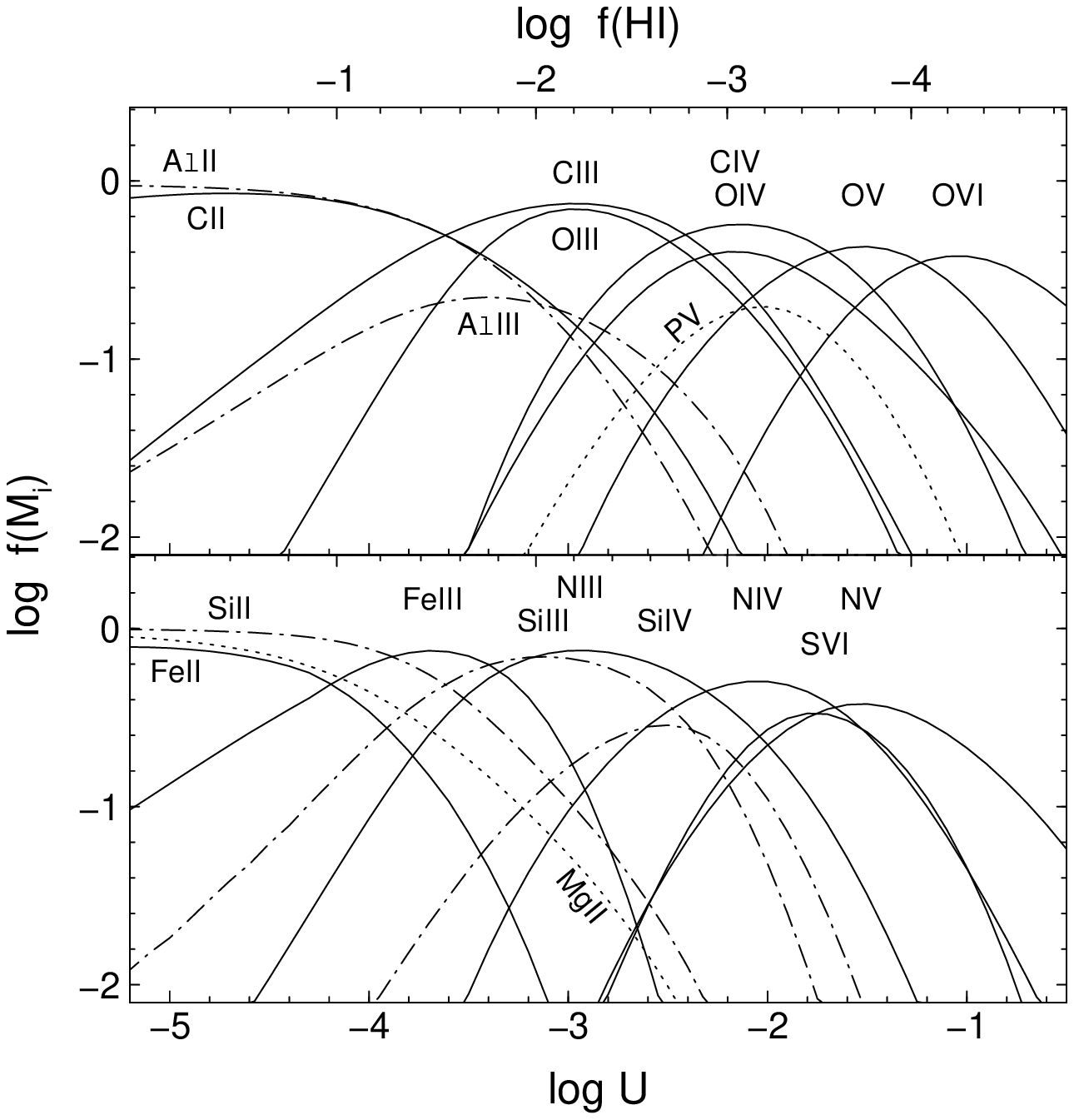}{4.6in}{0.0}{95.0}{95.0}{-293.0}{-294.0}
\end{figure}
\begin{quotation}
\noindent Fig. 10 --- Ionization fractions 
in optically thin clouds photoionized at different $U$ by a 
power-law spectrum with $\alpha =$~$-$1.5. The HI fraction 
appears across the top. The curves for the metal ions 
are labeled above or below their peaks whenever possible. The 
notation here is HI = H$^o$, CIV = C$^{+3}$, etc.
\end{quotation}

Estimates of $IC$ generally come from plots 
like Figure 10, which shows the ionization fractions of  
HI and various metal ions $M_i$ in photoionized clouds 
(see \S2.5 and Ferland \etal 1998 for general descriptions 
of the calculations). 
Ideally, we would compare column densities in different ions of 
the same element to obtain abundance-independent constraints 
on the ionization and thus $IC$. Otherwise, column densities 
in different elements can also constrain $IC$ with assumptions 
about the relative abundances. Note that the results in 
Figure 10 are not sensitive to the particular abundances used 
the calculations (in this case solar), so the figure is useful 
for general abundance/ionization estimates (see Hamann 1997 
--- hereafter H97). 
The model clouds are optically thin in the ionizing UV continuum, 
which means that gradients in the ionization are negligible 
across the cloud and the ionization fractions do not 
depend on the total column densities. This simplification 
appears to be appropriate for most intrinsic 
absorption line systems (based on their measured column densities), 
although shielding by many far-UV BALs might affect the ionization 
structure downstream in BALRs (Korista \etal 1996, 
Turnshek 1997, H97). Also, systems with 
low-ionization lines like FeII or MgII can be optically 
thick at the HI Lyman edge (Bergeron \& Stasi\'nska 
1986, Voit \etal 1993, Wampler, Chugai \& Petitjean 1995) and 
may require calculations with specific 
column densities that match the data. 

\subsubsection{Ionization Ambiguities}

The main theoretical uncertainties involve the shape of the 
ionizing spectrum, the frequent 
lack of ionization constraints (too few lines measured), 
and the possibility of inhomogeneous (multi-zone) absorbing 
media. H97 addressed these issues 
by calculating $IC$ values for a wide range of conditions 
in photoionized clouds. He noted that, whenever there is or might 
be a multi-zone with a range of ionizations, we can still 
make conservatively low estimates of the metal-to-hydrogen 
abundance ratios by assuming each metal line forms 
where that ion is most favored --- that is, at the peak of its 
ionization fraction $f(M_i)$ in Figure 10. We can also place 
firm lower limits on the metal-to-hydrogen ratios by 
adopting the minimum values of $IC$, which correspond to 
minima in the $f$(HI)/$f$(M$_i$) ratios (see also Bergeron \& 
Stasi\'nska 1986). The lower limits are robust, even though they 
come from 1-zone calculations, because different or additional 
zones can only mean that larger $IC$ values are appropriate for 
the data. Figure 11 plots several minimum metal-to-hydrogen $IC$s 
for optically thin clouds photoionized by different power-law 
spectra. The results in this plot simply get added to the 
logarithmic column density ratios (Eqn. 10) to derive minimum 
metallicities. Note that some important metal-to-metal ratios also 
have minimum $IC$ values, such as PV/CIV and FeII/MgII 
(Hamann 1998, Hamann \etal 1999b). 

\subsubsection{Column Densities and Partial Coverage}

The final critical issue is deriving accurate column 
densities from the absorption troughs. In the simplest case, 
the line optical depths are related to the observed intensities by,
\begin{equation}
I_{v}\ =\ I_o\,\exp{(-\tau_{v})}
\end{equation}
where $I_v$ and $I_o$ are the observed and intrinsic 
(unabsorbed) intensities, respectively, and $\tau_{v}$ is the 
line optical depth, at each velocity shift $v$. The 
column densities follow from the optical depths by,
\begin{equation}
N\ =\ {{m_ec}\over{\pi e^2 f\lambda_o}}\,\int\tau_v\ dv
\end{equation}
\newpage
\begin{figure}[h]
\plotfiddle{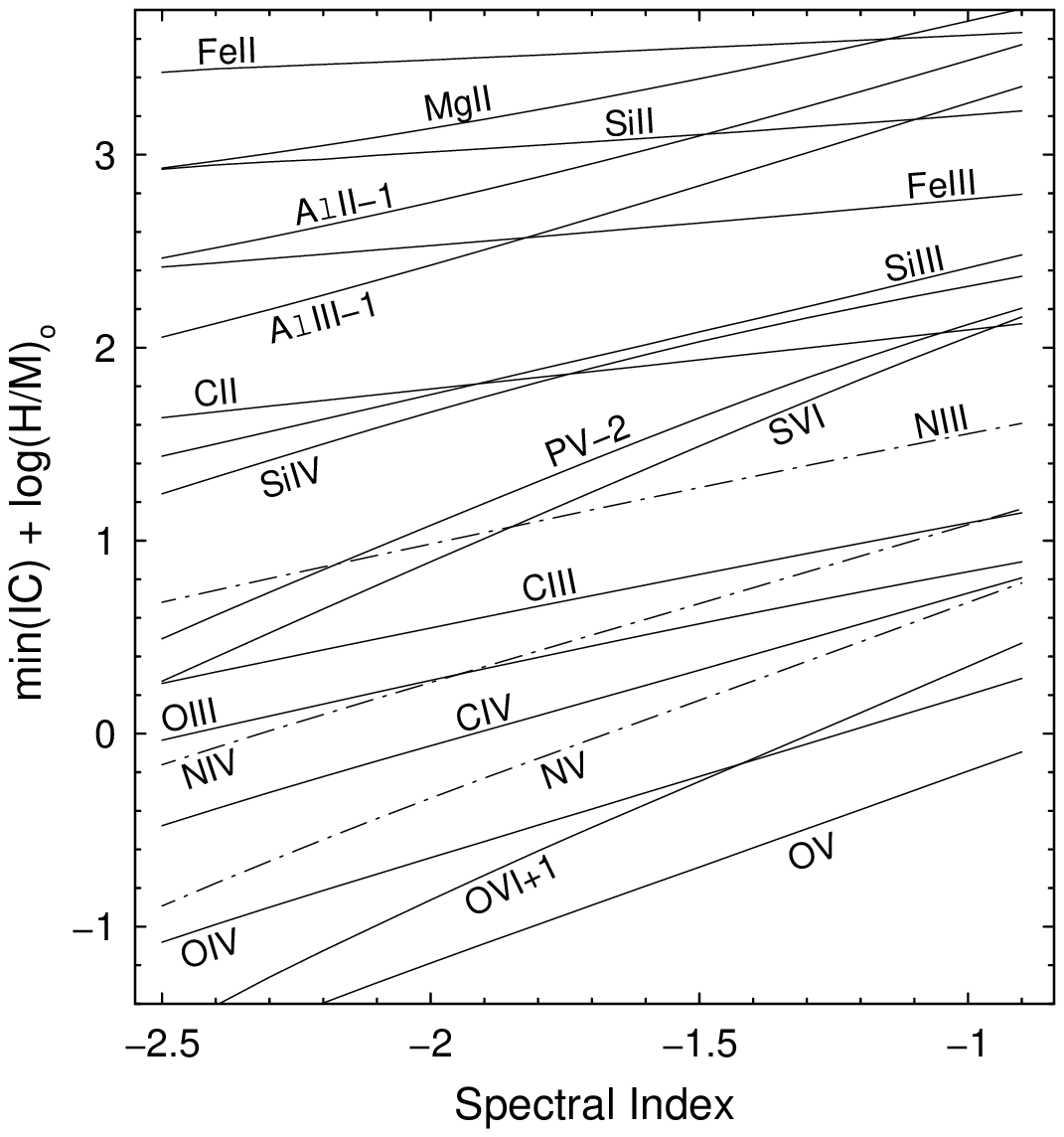}{4.0in}{0.0}{100.0}{100.0}{-320.0}{-273.0}
\end{figure}
\begin{quotation}
\noindent Fig. 11 --- Minimum metal ion to HI ionization 
corrections ($IC$) normalized to solar abundances (the last two 
terms in Eqn. 10) are plotted for optically thin clouds photoionized 
by power-law spectra with different indices ($\alpha$). The notation 
is the same as Fig. 10. The curves have been shifted vertically by 
+1 for OVI, by $-$2 for PV, and by $-$1 for AlII and AlIII. 
The curves for nitrogen ions are dash-dot.
\end{quotation}

\ni where $f$ is the oscillator strength and $\lambda_o$ is 
the laboratory wavelength of the line. 
Column density derivations can involve 
line profile fitting or direct integration over the observed 
profiles (via Eqns. 11 and 12, Junkkarinen \etal 1983, Grillmair 
\& Turnshek 1987, Korista \etal 1992, 
Savage \& Sembach 1991, Jenkins 1996, Arav \etal 1999 --- 
hereafter A99). Very optically 
thick lines are not useful because the inferred values of 
$\tau_{v}$ are far too sensitive to uncertainties 
in $I_{v}$. In other cases, 
the analysis might still be compromised by 1) unresolved 
absorption-line components or 2) unabsorbed flux that fills in 
the the bottoms of the observed troughs. If either of these 
possibilities occurs, the derived optical depths and 
column densities become lower limits and the derived 
abundances become incorrect. Errors from the first possibility 
can always be reduced or avoided by higher resolution spectroscopy.

\newpage
\begin{quotation}
\centerline{XXXXXX INSERT FIGURE HERE XXXXXX}
\noindent Fig. 12 --- 
Schematic showing possible ``partial coverage'' geometries. Partial 
line-of-sight coverage occurs when light rays like C, which pass through  
absorption-line clouds (indicated by filled ellipsoids), are combined  
with rays like A, B or D, which do not. Ray A represents reflected 
light from a putative scattering region. Ray B simply misses the 
absorption-line region. Ray D passes through the nominal absorbing 
zone but suffers no absorption because the region is porous.  
\end{quotation}

The second possibility, of filled-in absorption troughs, 
is actually an asset for identifying intrinsic NALs 
(\S3.1). We will refer to this filling-in generally as 
``partial coverage'' of the background emission source. 
Figure 12 shows several geometries that might produce 
partial coverage and filled-in troughs\footnote{The situation 
can be potentially more complicated if the absorber itself is 
a source of emission. The analysis discussed below remains the 
same, however.}. When partial coverage occurs, 
the observed intensities depend on both the optical depth 
and the line-of-sight coverage fraction, $C_f$, 
at each velocity, 
\begin{equation}
I_v\ =\ (1-C_f)\,I_o\, +\, C_f\,I_o\,\exp{(-\tau_{v})}
\end{equation}
where $0\leq C_f\leq 1$ and the first term on the right 
side is the unabsorbed (or uncovered) contribution. 
Measured absorption lines can thus be 
shallow even when the true optical 
depths are large. In the limit $\tau_{v}\gg 1$, we have, 
\begin{equation}
C_f\ =\ 1 - {{I_{v}}\over{I_o}}
\end{equation}
Outside of that limit, 
we can compare lines whose true optical depth ratios are fixed by 
atomic physics, such as the HI Lyman lines or doublets like 
CIV~\lam\lam 1548,1550, SiIV~\lam\lam 1394,1403, etc., to 
determine uniquely both the coverage fractions and the true 
optical depths across the line profiles 
(H97b, Barlow \& Sargent 1997, A99, Srianand \& 
Shankaranarayanan 1999, Ganguly \etal 1999). 
For example, a little algebra shows that for doublets with
true optical depth ratios of $\sim$2 (as in CIV, 
SiIV, etc.) the coverage fraction at each absorption velocity is, 
\begin{equation}
C_f\ =\ {{I_1^2 - 2I_1 + 1}\over{I_2 - 2I_1 +1}} 
\end{equation}
where $I_1$ and $I_2$ are the observed 
line intensities, normalized by $I_o$, at the same velocity in 
the weaker and stronger line troughs, respectively. 
The corresponding line optical depths are $\tau_2 = 2\tau_1$ and,
\begin{equation}
\tau_1\ =\ \ln\left({{C_f}\over{I_1 + C_f - 1}}\right)
\end{equation}

It is a major strength of the NALs that we can 
resolve key multiplet lines and use this analysis 
to measure the coverage fractions and thus derive reliable 
column densities and abundances. It is a great weakness 
of the BALs that this analysis is usually not 
possible because the lines are blended. We will 
argue below that BAL studies so far have been seriously 
compromised by unaccounted for partial-coverage effects.
 
The only drawback of partial coverage for the NALs is that 
there might be a range of coverage fractions in multi-zone 
absorbing media. There is already evidence in some cases 
for coverage fractions that differ between ions or 
change with velocity across the line profiles 
(Barlow \& Sargent 1997, Barlow, Hamann \& Sargent 1997, 
H97b). Variations in $C_f$ with velocity 
can always be dealt with by analyzing limited velocity intervals 
in the line profiles (see also Arav 1997). 
But one can imagine complex geometries where ionization-dependent 
coverage fractions would jeopardize the simple 
analysis described above, in particular for comparisons 
between high and low ionization species like CIV and HI. 
Abundance ratios based on disparate species like these might 
require specific models of the ionization-dependent coverage. 
On the other hand, this worst-case scenario is not known 
to occur, and there is no reason to believe it would lead 
to generally overestimated metallicities anyway. 

\subsection{Broad Absorption Line Results}

One common characteristic of BAL spectra is that the 
metallic resonance lines like CIV~\lam 1548,1951, SiIV~\lam 1394,1403, 
NV~\lam 1239,1243 and OVI~\lam 1032,1038 are typically  
strong (deep) compared to \Lya\ (e.g. Fig. 8). This result, 
and the fact that low-ionization lines like MgII~\lam 2796,2804 
and FeII (UV) are usually absent, indicates that the BALR ionization 
is generally high (Turnshek 1984, Weymann \etal 1981,1985). 
However, quantitative studies of the ionization have repeatedly 
failed to explain the measured line strengths with solar 
abundances. These difficulties were first noted by 
Junkkarinen (1980) and Turnshek (1981, also 
Weymann \& Foltz 1983), who  
showed that photoionization models with  
power-law ionizing spectra and solar abundances 
underpredict the metal ions, especially SiIV, by large 
factors relative to HI. A straightforward 
conclusion is that the metallicities are 
well above solar. Turnshek (1986, 1988) and 
Turnshek \etal (1987) estimated 
metal abundances (C/H) of 10 to 100 times solar, 
and provided tentative evidence for some extreme 
metal-to-metal abundance ratios such as P/C~$\ga$~100 
times solar. 

Better data in the past ten years have 
done nothing to change these startling results 
(e.g. Turnshek \etal 1996). The early concerns about 
unresolved line components (Junkkarinen \etal 1987, Kwan 1990) 
have gone away, thanks to spectroscopy with 
the Keck 10 m telescope at resolutions ($\sim$7~\kms ) close 
to the thermal speeds (Barlow \& Junkkarinen 1994, 
Junkkarinen 1998). The previously tentative detections of 
PV~\lam 1118,1128 absorption, which led to the large 
P/C abundance estimates, have now been confirmed in 
two objects by excellent wavelength coincidences, by 
the predicted weakness of nearby lines like FeIII~\lam 1122, 
and in one case by the probable presence of PIV~\lam 951 
absorption (Junkkarinen \etal 1997, Hamann 1998, Fig. 8). 
The commonality of PV absorption is not yet known 
(see also Korista \etal 1992, Turnshek \etal 1996), but its 
relative strength in just the two cases is surprising because 
the solar P/C ratio is only $\sim$0.001. 

More complex theoretical analyses, considering  
a range of ionizing spectral shapes or 
multiple ionization zones, also do not change the main result 
for metallicities and P/C ratios well above solar 
(Weymann \etal 1985,  Turnshek \etal 1987, 1996 and 1997, 
Korista \etal 1996). H97 used the analysis 
in \S3.2 to determine how high the abundances 
must be given the measured column densities and a 
photoionized BALR. He showed that average BALR column densities 
require [C/H] and [N/H]~$>$~0 and [Si/H]~$>$~1.0 for any range of 
ionizations and reasonable spectral shapes. The conservatively 
low (but not quite minimum) values of $IC$ indicate 
[C/H] and [N/H]~$\ga$~1.0 and [Si/H]~$\ga$~1.7.  
The results for individual BAL 
systems can be much higher. In PG1254+047 (Fig. 8, Hamann 1998) 
the inferred minimum abundances are [C/H] and [N/H]~$\ga$~1.0, 
[Si/H]~$\ga$~1.8 and [P/C]~$\ga$~2.2. 

However, we will now argue that all these BAL abundance 
results are incorrect, because 
partial coverage effects have led to generally underestimated 
column densities.

\subsubsection{Uncertainties and Conclusions}

There is now direct evidence for partial coverage in some BALQSOs 
based on widely separated lines of the same ion (A99) and resolved 
doublets in several narrow BALs and BAL components 
(Telfer \etal 1999, Barlow \& Junkkarinen 1994, Wampler \etal 1995, 
Korista \etal 1992 --- confirmed by Junkkarinen 1998). 
Although most this evidence applies to narrow features, 
it is noteworthy that there are no 
counterexamples to our knowledge --- where narrow line 
components associated with BALs indicate complete coverage 
(also Junkkarinen 1998). 

There is also circumstantial evidence for partial coverage 
in BAL systems. Namely, 1) 
spectropolarimetry indicates that BAL troughs can be filled in by 
polarized flux (probably from an extended scattering 
region) that is not covered by the BALR (Fig. 12, Goodrich \& Miller 
1995, Cohen \etal 1995, Hines \& Wills 1995, Schmidt \& Hines 1999). 
2) Some BAL systems have a wide range of lines with suspiciously 
similar strengths or flat-bottom troughs that do not reach zero
intensity (Arav 1997). 3) Voit \etal (1993) made 
a strong case for low-ionization BALRs 
being optically thick at the Lyman limit, 
which implies large optical depths in \Lya , yet the \Lya\ 
troughs are not generally black in these systems.
4) The larger column densities that follow assuming partial 
coverage and saturated BALs 
($N_H\ga 10^{22}$~\cmsq , Hamann 1998) are consistent with 
the large absorbing columns inferred from X-ray observations 
of BALQSOs (Green \& Mathur 1996, Green \etal 1997, 
Gallagher \etal 1999). 

More indirect evidence comes from the abundance results 
themselves. Voit (1997) noted that the derived overabundances 
tend to be greater for rare elements like P than for 
common elements like C. This is precisely what would occur 
if line saturation is not taken into account. The 
surprising detections of PV might actually be a signature of 
line saturation (and partial coverage) in strong lines like CIV, 
rather than extreme abundances (Hamann 1998). 
This assertion is supported by the one known NAL system with 
PV~\lam\lam 1118,1128 absorption, where  
the doublet ratios in CIV, NV and SiIV clearly indicate 
$\tau\gg$1 (Barlow \etal 1997, Barlow 1998). 

We conclude that BAL column densities have been generally 
underestimated and the true BALR abundances are not 
known. Observed differences between BAL profiles that resemble 
simple optical depth effects are probably caused by a mixture 
of ionization, coverage fraction and optical depth differences 
in complex, multi-zone BALRs. 
This conclusion paints a grim picture for BAL abundance work, 
but it might still be possible to derive accurate 
column densities and therefore abundances for some BALQSOs or 
some portions of BAL profiles (Wampler \etal 1995, Turnshek 1997, 
A99). Most needed are spectra at shorter rest-frame 
wavelengths to measure widely separated lines of the same ion 
and thereby diagnose the coverage fractions and true optical depths 
(\S3.2.2, Arav 1997, A99). 

\subsection{Narrow Absorption Line Results}

In contrast to the BALs, 
intrinsic NALs might be the best abundance probes we 
have for QSO environments. Resolved measurements of NAL 
multiplets allow us to measure both the coverage fractions 
and true column densities (\S3.2.2). The NALs also 
allow separate measurements of important lines that are 
often blended in BAL systems, such as NV~\lam 1239,1243--\Lya ,  
OVI~\lam 1032,1038--\Lyb\ and many others. 
We therefore have potentially many more constraints on 
both the ionization and abundances. 

Early NAL studies did not have the quality of data needed 
to derive column densties and abundances, but several 
groups noted a tendency for larger NV/CIV  
line strength ratios in \zaz\ systems compared to \zllz\ 
(Weymann \etal 1981, Hartquist \& Snijders 1982, 
Bergeron \& Kunth 1983, Morris \etal 1986, 
Bergeron \& Boiss\'e 1986). This trend is probably 
not due simply to higher ionization in \zaz\ absorbers, 
because recent studies show that \zllz\ 
systems typically have strong OVI lines and 
therefore considerable high-ionization gas; NV appears 
to be weak relative to both CIV and OVI 
at \zllz\ (Lu \& Savage 1993, Bergeron \etal 1994, 
Burles \& Tytler 1996, Kirkman \& Tytler 1997, 
Savage, Tripp \& Lu 1998). The lower NV/OVI and NV/CIV line 
ratios at \zllz\ could be caused by an underabundance of 
nitrogen (relative to solar ratios) in metal-poor intervening 
gas (Bergeron \etal 1994, Hamann \etal 1997d, 
Kirkman \& Tytler 1997). This would be the 
classic abundance pattern involving secondary nitrogen 
(Vila-Costas \& Edmunds 1993). 
Relatively higher N abundances and thus stronger NV 
absorption lines should occur naturally in metal-rich 
environments near QSOs (see \S\S6--7 below).  

The first explicit estimates of \zaz\ metallicities 
were by Wampler \etal (1993), M\"oller \etal (1994), Petitjean 
\etal (1994) and Savaglio \etal (1994) for QSOs at redshifts of 
roughly 2 to 4. These studies found that \zaz\ systems often 
have $Z\ga$~\Zsun , which is at least an order of magnitude 
larger than the \zllz\ systems measured in the same 
data. Several of the metal-rich \zaz\ systems have doublet 
ratios implying partial coverage and thus, very likely, an intrinsic 
origin (Wampler \etal 1993, Petitjean \etal 1994). 
The location of the other \zaz\ absorbers is not known,  
but Petitjean \etal (1994) noted a marked change from 
[C/H]~$\la$~$-$1 to [C/H]~$\ga$~0 at a blueshift of 
$\sim$15,000~\kms\ relative to the emission lines. If high 
abundances occur only in intrinsic systems, then these results 
suggest that most \zaz\ NALs are intrinsic (also M\"oller 
\etal 1994). 

More recent studies support these findings. 
Petitjean \& Srianand (1999) measured $Z\ga$~\Zsun\ and 
[N/C]~$>0$ in an intrinsic (partial coverage) \zaz\ absorber. 
For \zaz\ systems of unknown origin, 
Savage \etal (1998) estimated roughly solar metallicities 
and Tripp \etal (1997) obtained [N/C]~$\ga 0.1$ and, very 
conservatively, [C/H]~$\ga -0.8$. (The lower limit on [C/H] 
for the latter system is $-$0.2 when more likely ionizing 
spectral shapes are used in the calculations.)  
Savaglio \etal (1997) revised the 
metallicities downward slightly from their 1994 paper to  
$-1<$~[C/H]~$<0$, based on better data. Those systems 
are of special interest because of their  
high redshift ($z_a\approx 4.1$). 
Wampler \etal (1996) estimated $Z\sim 2$~\Zsun\ (based 
on a tentative detection of OI~\lam 1303) 
for the only other \zaz\ systems studied so far at $z>4$. 

H97, H97b and Hamann \etal (1995, 1997e, 1999b) 
used the analysis outlined in \S3.2 to 
determine metallicities or establish lower limits 
for several \zaz\ systems, including some mentioned above and 
some that are clearly intrinsic by the indicators in \S3.1. 
The results generally confirm the previous estimates 
and show further that, even when there are 
no constraints on the ionization (for example, when only 
\Lya\ and CIV lines are measured), the column densities can 
still require $Z\ga$~\Zsun . A quick survey of those results 
suggests  that bona fide intrinsic systems, 
and most others with $Z\ga$~\Zsun , have [N/C]~$\ga 0.0$.

\subsubsection{Uncertainties and Conclusions}

Most of the studies mentioned above would benefit from 
better data (higher signal-to-noise ratios and higher 
spectral resolutions) and more ionization constraints 
(wider wavelength coverage), 
but the frequent result for $Z\ga$~\Zsun\ is convincing. 
Unlike the BALs, there are no obvious systematic effects 
that might lead to higher abundance estimates for \zaz\ 
systems compared to \zllz . The possibility of ionization-dependent 
coverage fractions presents an uncertainty for those systems 
with partial coverage, but we do not expect that to cause 
systematic overestimates of the metallicities (\S3.2.2). 
We conclude that many \zaz\ NALs and, more importantly, all 
of the ``confirmed'' intrinsic systems, have $Z\ga$~0.5\Zsun\ 
and usually $Z\ga$~\Zsun . The upper limits on $Z$ are 
uncertain. The largest estimate for a well-measured system 
is $Z\sim 10$~\Zsun\ (Petitjean \etal 1994), 
but those data are also consistent with metallicities as low as 
solar because of ionization uncertainties (H97). 
There are mixed and confusing reports in the literature regarding 
metal-to-metal abundance ratios, most notably N/C. In contrast 
to Franceschini \& Gratton (1997), we find no tendency for 
sub-solar N/C in \zaz\ systems. In fact, there is the general 
trend for stronger NV absorption at \zaz\ compared to \zllz\ 
systems, and the most reliable abundance data suggest solar or 
higher N/C ratios whenever $Z\ga$~\Zsun . 

The only serious problem is in interpreting the abundance 
results for systems of unknown origin. High metallicities 
might correlate strongly with absorption near QSOs, 
but the metallicities cannot define the absorber's location. 
For example, Tripp \etal (1996) estimated $Z\ga$~\Zsun\ and 
[N/C]~$\ga$~0 for a \zaz\ system where the lack 
of excited-state absorption in CII$^*$~\lam 1336 
(compared the measured CII~\lam 1335) 
implies that the density is low, $\la$7 \pcc , and thus 
the distance from the QSO is large, $\ga$300~kpc. 
(The relationship between density and distance follows 
from the flux requirements for photoionization, 
\S2.5.1.) Super-solar metallicities at these large distances 
are surprising. At $\ga$300~kpc from the QSO, we might have 
expected very low intergalactic or halo-like abundances. 
The solution might be that the absorbing gas was enriched much 
nearer the QSO and then ejected (Tripp \etal 1996). 

Unfortunately, the excited-state lines used for density and 
distance estimates are not generally available for \zaz\ systems 
(because they have low ionization energies, e.g. CII$^*$ and 
SiII$^*$). Of the six \zaz\ absorbers known to be far ($\ga$10~kpc) 
from QSOs based on these indicators, 
three of them clearly have \zgz\ and are probably not intrinsic 
for that reason (Williams \etal 1975, Williams \& Weymann 1976, 
Sargent \etal 1982, Morris \etal 1986, 
Barlow \etal 1997). Only one has a 
metallicity estimate --- the system with $Z\ga$~\Zsun\ at 
$\ga$300~kpc distance studied by Tripp \etal (1996). 

\section{General Abundance Summary}

The main abundance results are as follows. 

1) There is a growing consensus from the BELs and 
NALs for $Z\ga$~\Zsun\ in QSOs out to $z>4$. 
The upper limits on the metallicities are not well known, 
but none of the data require $Z>10$~\Zsun . 
Solar to a few times solar appears to be typical. Based 
on very limited data, there is no evidence for a decline at 
the highest redshifts.

2) A trend in the NV/HeII and NV/CIV BEL ratios suggests that 
the metallicities are generally higher in more luminous QSOs. 

3) The BELs and NALs both suggest that the relative nitrogen 
abundance (e.g. N/C and N/O) is typically solar or higher. 
We will argue below (\S6) that this result corroborates the 
evidence for $Z\ga$~\Zsun\ (because of the likely secondary 
origin of nitrogen at these metallicities).

4) There is tentative evidence for super-solar Fe/Mg abundances 
out to $z>4$ based on the FeII/MgII BEL strengths. Again, 
based on limited data, there is no evidence for a decline in this 
ratio at the highest redshifts.

5) The extremely high metallicities and large P/C ratios 
derived so far from the BALs 
are probably incorrect. In further support of that conclusion, 
we note that BELR simulations using the nominally derived 
BAL abundances (including large enhancements in P and 
other odd-numbered elements like Al, Shields 1996) are 
inconsistent with observed BEL spectra (based on unpublished 
work in collaboration with G. Shields). 

\section{Enrichment Scenarios}

Several scenarios have been proposed for the production of 
heavy elements near QSOs, including 1) the normal evolution 
of stellar populations in galactic nuclei (Hamann \& Ferland 1992, 
HF93b), 2) central star clusters with enhanced supernova (and 
perhaps nova) rates due to mass accreted onto stars as they plunge 
through QSO accretion disks (Artimowitz, Lin \& Wampler 1993), 
3) star formation inside QSO accretion disks (Silk \& Rees 1998, 
Collin 1998), and 4) nucleosynthesis without stars inside 
accretion disks (Jin, Arnett \& Chakrabarti 1989, Kundt 1996). 

\subsection{Occam's Razor: The Case for Normal Galactic Chemical 
Evolution}

The first scenario listed above, for normal galactic 
chemical evolution, is most compelling because 1) it is the only 
one of these processes known to occur and 2) it is sufficient to 
explain the QSO data. In particular, the stars in the centers of 
massive galaxies today are (mostly) old and metal rich 
(Bica, Arimoto \& Alloin 1988, 
Bica, Alloin \& Schmidt 1990, 
Gorgas, Efstathiou \& Arag\' on Salamanca 1990, Bruzual \etal 
1997, Vazdekis \etal 1997, Jablonka, Alloin \& Bica 1992, 
Jablonka, Martin \& Arimoto 1996, 
Feltzing \& Gilmore 1998, Worthy, Faber \& Gonzalez 1992, 
Kuntschner \& Davies 1997, 
Sansom \& Proctor 1998, Ortolani \etal 1996, 
Sil'chenko, Burenkov \& Vlasyuk 1998, 
Idiart, de Freitas Pacheco, \& Costa 1996, 
Fisher, Franx \& Illingworth 1995, Bressan, Chiosi \& Tantalo 1996). 
The exact ages are uncertain, but there is growing evidence for most 
of the star formation in massive spheroids 
(ellipticals and the bulges of large spiral galaxies) occurring 
at redshifts $z\ga2$--3, especially (but not only) for galaxies 
in clusters (see also Renzini 1998,1997, Bernardi \etal 1998, 
Bruzual \& Magris 1997, Ellis \etal 1997, Tantalo, Chiosi 
\& Bressan 1998, Ivison \etal 1998, Kodama \& Arimoto 1997, 
Ziegler \& Bender 1997, Kauffmann 1996, 
Van Dokkum \etal 1998, Mushotzky \& Loewenstein 1997
Spinrad \etal 1997, Stanford \etal 1998, Heap \etal 1998, 
Barger \etal 1998a,b).  The star-forming (Lyman-break or 
\Lya -emission) objects 
measured directly at $z\ga 3$ might be galactic or proto-galactic 
nuclei in the throes of rapid evolution (Friaca \& Terlevich 1999, 
Baugh \etal 1998, 
Steidel \etal 1998 and 1999, Connolly \etal 1997, Lowenthal \etal 1997, 
Trager \etal 1997, Hu, Cowie \& McMahon 1998, 
Franx \etal 1997, Madau \etal 1996, Giavalisco, Steidel \& Machetto 
1996). These objects are more numerous than QSOs and some 
have been measured at $z>5$ (Dey \etal 1998, 
Hu \etal 1998, Weymann \etal 1998), beyond the highest known 
QSO redshift of $z\approx 5.0$ (Sloan Digital Sky Survey 
press release 1998). On the theoretical 
side, recent cosmic-structure simulations show that proto-galactic 
condensations can form stars and reach solar or higher 
metallicities at $z\ga 6$ (Gnedin \& Ostriker 1997).
Quasars might form in the most massive and most dense 
of these early-epoch star-forming environments 
(Turner 1991, Loeb 1993, Haehnelt \& Rees 1993, Miralda-Escude \& 
Rees 1997, Haehnelt \etal 1998, Spaans \& Carollo 1997). They might 
also form preferentially in globally dense cluster environments, 
based on the higher detection rates of star-forming galaxies near 
high-$z$ QSOs (Djorgovski 1998). 

The gas in these environments might have been long 
ago ejected via galactic winds, consumed by central black holes 
or diluted by subsequent infall, but its signature 
remains in the old stars today. The mean stellar 
metallicities\footnote{It is worth noting here that, because 
of a significant time-delay in the iron enrichment,  
O/H and Mg/H are better measures of the overall ``metallicity'' 
than Fe/H (see \S6 and Wheeler \etal 1989).} 
in the cores of massive low-redshift galaxies are 
typically $\langle Z_{stars}\rangle \sim 1$--3~\Zsun\ 
(see refs. listed above). 
Individual stars are distributed about the 
mean with metallicities reflecting the gas-phase abundance 
at the time of their formation. If the interstellar gas is well-mixed 
and the abundances grow monotonically (as expected in simple 
enrichment schemes, \S6), the gas-phase metallicity, $Z_{gas}$, 
will always exceed $\langle Z_{stars}\rangle$. Only the most 
recently formed stars will have metallicities as high as the gas. 
Therefore, the most metal-rich stars today should reveal the 
gas-phase abundances near the end of the last major 
star-forming epoch. 

In the bulge of our own Galaxy, the nominal value of 
$\langle Z_{stars}\rangle$ is 1~\Zsun\ and the tail of the 
distribution reaches $Z_{stars}\ga 3$~\Zsun , with even higher 
values obtaining near the Galactic center (Rich 1988 and 1990, 
Geisler \& Friel 1992, McWilliam \& Rich 1994, Minniti \etal 
1995, Tiede, Frogel \& Terndrup 1995, Terndrup, Sadler \& Rich 
1995, Idiart \etal 1996, Castro \etal 1996, Bruzual \etal 1997). 
The gas-phase metallicity should therefore have been 
$Z_{gas}\ga 3$~\Zsun\ after most of the Bulge star formation 
occurred. Simple chemical evolution models 
indicate more generally that $Z_{gas}$ should be $\sim$2 
to 3 times $\langle Z_{stars}\rangle$ in spheroidal systems 
like galactic nuclei (Searle \& Zinn 1978, Tinsley 1980, Rich 
1990, Edmunds 1992, de Fretas Pacheco 1996). Thus the observations 
of $\langle Z_{stars}\rangle \sim 1$--3~\Zsun\ suggest that gas 
with $Z_{gas}\sim 2$--9~Z$_{\odot}$ once existed in these 
environments. 

We might therefore expect to find 
$2\la Z\la 9$~\Zsun\ in QSOs, as long as 1) most of the 
local star formation occurred before the QSOs ``turned on'' or 
became observable and 2) the metal-rich gas produced by that 
star formation was not substantially diluted or ejected. 
These expectations are consistent with the abundance estimates 
reported above (\S4). More exotic enrichment schemes 
are therefore not needed to explain the QSO data. 

\section{More Insights from Galactic Chemical Evolution}

If we assume that QSO environments were indeed enriched by 
normal stellar populations, then we can use the results from galactic 
abundance and chemical evolution studies to interpret the QSO data. 
Here we describe some relevant galactic results (see Wheeler, 
Sneden \& Truran 1989 for a general review). 

\subsection{The Galactic Mass-Metallicity Relation}

One important result from galaxy studies  
is the well-known mass--metallicity 
relationship among ellipticals and spiral bulges (Faber 1973,  
Faber \etal 1989, 
Bender, Burstein \& Faber 1993, Zaritsky, Kennicutt \& Huchra 1994, 
Jablonka \etal 1996, Coziol \etal 1997). This relationship is 
attributed to the action of galactic winds;  
massive galaxies reach higher metallicities because they have  
deeper gravitational potentials and are better able to retain 
their gas against the building thermal pressures 
from supernovae (Larson 1974, 
Arimoto \& Yoshii 1987, Franx \& Illingworth 1990). Low-mass 
systems eject their gas before high $Z$'s are attained. 
Quasar metallicities should be similarly tied to 
the gravitational binding energy of the local star-forming 
regions and, perhaps, to the total masses of their host galaxies 
(\S7.1 below). 

\subsection{Specific Abundance Predictions}

Another key result is the abundance behaviors of N and 
Fe relative to the $\alpha$ elements such as O, Mg and Si. 
HF93b constructed 1-zone infall models of galactic chemical 
evolution to illustrate these behaviors in different 
environments. Figure 13 plots the results for two scenarios 
at opposite extremes. Both use the same nucleosynthetic 
yields, but the ``Giant Elliptical'' model has much faster evolution 
rates and a flatter IMF (more favorable to high-mass stars) 
compared to the ``Solar Neighborhood'' (or spiral disk) case. 
The Giant Elliptical evolves passively (without further star 
formation) after $\sim$1~Gyr because the gas is essentially 
exhausted. The parameters used in these calculations were 
based on standard galactic infall models (e.g. 
Arimoto \& Yoshii 1987, Matteucci \& Tornamb\'e 1987, 
Matteucci \& Francois 1989, Matteucci \& Brocato 1990, 
K\"oppen \& Arimoto 1990). However, the results are only 
illustrative and not intended to match entire galaxies. 
For example, evolution like the Giant Elliptical model might 
occur in just the central cores of extreme high-mass 
galaxies (cf. Friaca \& Terlevich 1998).

\begin{figure}[h]
\plotfiddle{figure13.ps}{2.3in}{0.0}{50.0}{50.0}{-200.0}{-475.0}
\end{figure}
\begin{quotation}
\noindent Fig. 13 --- Logarithmic gas-phase abundance ratios 
normalized to solar for the two evolution models 
discussed in \S6.2 (adapted from HF93b). Two scenarios 
for the N enrichment are shown (thin solid lines); one 
with secondary only and the other with secondary+primary 
(causing a plateau in N/O at low $Z$). 
\end{quotation}

\subsection{Fe/$\alpha$ as a Clock}

At early times the abundance evolution 
is controlled by short-lived massive 
stars, mainly via type II supernovae (SN~II's). 
The $\alpha$ elements, such as O and Mg, come almost 
exclusively from these objects, but Fe has a large delayed 
contribution from type Ia supernovae (SN~Ia's) --- whose 
precursors are believed to be intermediate mass stars in close 
binaries (Branch 1998). The predicted time delay is roughly 
1~Gyr based on the IMF-weighted stellar lifetimes (Fig. 13, 
Greggio \& Renzini 1983, Matteucci \& Greggio 1986). 
The actual delay is uncertain, but recent estimates 
are in the range $\sim$0.3 to 3~Gyr (Matteucci 1994, Yoshii, 
Tsujimoto \& Nomoto 1996, Yoshii \etal 1998). 
Because this delay does not depend on any of the global 
evolution time scales (e.g. the star formation rate, etc.), 
Fe/$\alpha$ can serve as an absolute ``clock'' for 
constraining the ages of star-forming environments (Tinsley 1979, 
Thomas, Greggio \& Bender 1998). 

Observations of 
metal-poor Galactic stars suggest that the baseline value of 
[Fe/$\alpha$] due to SN~II's alone is nominally 
$-$0.7 to $-$0.4 (Israelian, Garcia 
\& Rebolo 1998, Nissen \etal 1994, King 1993, 
Gratton 1991, Magain 1989, Barbuy 1988, also 
de Freitas Pacheco 1996), 
which is slightly larger than the prediction in Figure 13. 
The subsequent increase caused by SN~Ia's is a factor of a few 
or more. Note that the increase in Fe/$\alpha$ should be 
larger in rapidly-evolving spheroidal systems because 1) by the 
time their SN~Ia's ``turn on,'' there is relatively little gas 
left and each SN~Ia has a greater effect, also 2) their rapid early 
star formation means that the SN~Ia's occurring later are more 
nearly synchronized. The net result can 
be substantially super-solar Fe/$\alpha$ in the gas (even 
though Fe/$\alpha$ is sub-solar in most stars). 

\subsection{Nitrogen Abundances}

Nitrogen also exhibits a delayed enhancement, although not 
on a fixed time scale like Fe/$\alpha$. Nitrogen's selective 
behavior is due to secondary CNO nucleosynthesis, where N 
forms out of pre-existing C and O. Studies 
of galactic HII regions indicate that secondary processing 
dominates at metallicities above $\sim$0.2 \Zsun , 
resulting in N/O scaling like O/H (or N~$\propto Z^2$) in that 
regime. At lower metallicities, primary N can be more 
important based on an observed plateau in [N/O] at roughly $-$0.7
(see Tinsley 1980, Vila-Costas \& Edmunds 1993, Thurston, Edmunds 
\& Henry 1996, Van Zee, Skillman \& Salzer 1998, 
Kobulnicky \& Skillman 1998, Thuan, Izotov \& Lipovetsky 1995, 
Izotov \& Thuan 1999, but see also Garnett 1990, Lu \etal 1998). 
The models in Figure 13 show two N/O behaviors, for secondary 
only and secondary+primary, where the latter has a low-$Z$ plateau 
forced to match the HII region data. Notice that the secondary growth 
in N/O can be shifted down considerably from the simple theoretical 
relation [N/O]~=~[O/H] because of the delays related to stellar 
lifetimes.  We therefore have a strong prediction, based on both 
observations and these simulations, that measured values of 
[N/O]~$\ga$~0 imply $Z\ga$~\Zsun\ --- especially in fast-evolving 
spheroidal systems. This prediction was exploited above in 
the analysis of QSO BELs (\S2.6, Shields 1976). 

\section{Implications of QSO Abundances}

\subsection{High-Redshift Star Formation}

We can conclude from the previous sections that QSOs are 
associated with vigorous star formation, 
consistent with the early-epoch evolution of 
massive galactic nuclei or dense proto-galactic clumps (\S5). 
However, QSO abundances provide new constraints. 
For example, the general result for $Z\ga$~\Zsun\ 
suggests that most of the enrichment and local star formation 
occurs before QSOs ``turn on'' or become observable. 
The enrichment times can be 
so short in principle (Fig. 13, HF93b) that the star 
formation might also be coeval with QSO formation. In any 
event, the enrichment times cannot be much longer that 
$\sim$1~Gyr for at least the highest redshift objects 
(depending on the cosmology, Fig. 1).

If the QSO metallicities representative of a well-mixed 
interstellar medium, we can conclude further that the star 
formation was extensive. That is, a significant fraction 
of the initial gas must be converted into stars and stellar 
remnants to achieve $Z_{gas}\ga$~\Zsun . The exact fraction 
depends on the IMF. A solar neighborhood IMF (Scalo 1990, 
as in the Solar Neighborhood model of \S6) would lead to 
mass fractions in gas of only $\la$15\% at $Z\sim$~\Zsun , 
and would not be able to produce $Z_{gas}$ above a few \Zsun\ 
at all. Flatter IMFs (favoring massive stars) could reach 
$Z_{gas}\ga$~\Zsun\ while consuming less of the gas. 
For example, the gas fraction corresponding to $Z_{gas}\sim$~\Zsun\ 
in the Giant Elliptical model of \S6.2 is nearly 70\%. 

Figure 7 in \S2.6.3 illustrates the main star formation characteristics 
required by the QSO data. The solid curves on the right-hand side 
of that figure show theoretical BEL ratios from photoionization simulations 
that use nominal BELR parameters and abundances from the two 
chemical evolution models in Figure 13 (see HF93b for more details). 
The evolution is assumed to 
begin with the Big Bang and the conversion of time into redshift 
assumes a cosmology with $H_o = 65$~km~s$^{-1}$~Mpc$^{-1}$, 
$\Omega_{M} = 1$ and $\Omega_{\Lambda} = 0$ (Fig. 1). 
The main results are that the 
Solar Neighborhood evolution is too slow and, in any case, 
does not reach high enough metallicities or nitrogen enhancements 
to match most of the high-redshift QSOs. Much shorter time 
scales and usually higher metallicities, as in the Giant Elliptical 
simulation, are needed. 

A trend in the NV BELs suggests further that the 
metallicities are typically higher in more luminous QSOs 
(\S2.6.3). That result needs confirmation, but it could 
result naturally from a mass--metallicity relationship 
among QSO host galaxies that is similar (or identical to) 
the well-known relation in low-redshift galaxies (\S6, HF93b). 
By analogy with the galactic relation, the most luminous 
and metal-rich QSOs might reside in the most dense 
or massive host environments. This situation would be 
consistent with studies showing that QSO luminosities, 
QSO masses, and central black hole masses in galactic nuclei 
all appear to correlate with the mass of the 
surrounding galaxies (McLeod \etal 1999, McLeod \& Rieke 1995, 
Bahcall \etal 1997, Magorrian \etal 1998, Laor 1998, 
also Haehnelt \& Rees 1993). A direct application of 
the galactic mass-metallicity relation suggests 
that metal-rich QSOs reside in galaxies 
(or proto-galaxies) that are minimally as massive 
(or as tightly bound) as our own Milky Way. 

\subsection{Fe/$\alpha$: Timescales and Cosmology}

One of the most interesting predictions from galactic  
studies (\S6) is that Fe/$\alpha$ ratios in QSOs 
might constrain the epoch of their first star 
formation and perhaps the cosmology. In particular, 
large Fe/$\alpha$ ratios (solar or higher) would suggest 
that the local stellar populations are at least 
$\sim$1~Gyr old. At the highest QSO redshifts ($z\sim 5$), this age 
constraint would push the epoch of first star formation beyond 
the limits of current direct observation, 
to $z>6$ (Fig. 1). The $\sim$1~Gyr constraint 
would also be difficult to 
reconcile with $\Omega_M\approx 1$ in Big Bang cosmologies. 
Conversely, measurements of low Fe/$\alpha$ would suggest that 
the local stellar populations are younger than $\sim$1~Gyr  
(although we could not rule out the possibility that only 
SN~II's contributed to the enrichment for some reason). 
Some BEL studies have already suggested that Fe/$\alpha$ is 
above solar in $z>4$ QSOs (\S2.6.4).  

\subsection{Comparisons to Other Results}

Quasar abundances should be viewed in the context of other 
measures of the metallicity and star formation at high redshifts. 
Damped-\Lya\ absorbers in QSO spectra, 
which probe lines of sight through large intervening galaxies 
(probably spiral disks, Prochaska \& Wolfe 1998), 
have mean (gas-phase) metallicities of order 0.05~\Zsun\ at 
$z\ga 2$ (Lu \etal 1996, Pettini \etal 1997, 
Lu, Sargent \& Barlow 1998, Prochaska \& Wolfe 1999). 
The \Lya\ forest absorbers, which 
presumably probe much more extended and tenuous inter-galactic 
structures (Rauch 1998), typically have metalicities $<$0.01~\Zsun\ 
at high redshifts (Rauch, Haehnelt \& Steinmetz 1997, 
Songalia \& Cowie 1996, Tytler \etal 1995). 
The much higher metal abundances near QSOs are consistent 
with the rapid and more extensive evolution 
expected in dense environments (Gnedin \& Ostriker 1997). 
Perhaps this evolution is similar to that occurring in the many 
star-forming objects that are now measured directly 
at redshifts comparable to, and greater than, the QSOs (see 
refs. in \S5.1). 

The detections of strong dust and molecular gas emissions from 
QSOs support the evidence from their high abundances that 
considerable local star formation preceded the QSO epoch. The 
dust and molecules, presumably manufactured by stars, appear 
even in QSOs at $z\ga4$ (Isaac \etal 1994, 
Omont \etal 1996, Guilloteau \etal 1997). 

\section{Future Prospects}

We now have the observational and theoretical abilities 
to test and dramatically extend all of the QSO abundance work 
discussed above. The most pressing needs are to 
1) develop more independent abundance 
diagnostics, and 2) obtain more and better data to 
compare diagnostics in 
large QSO samples --- spanning a range of redshifts, luminosities, 
radio properties, etc. Absorption line studies 
will benefit generally from higher spectral resolutions and wider 
wavelength coverage, providing more accurate column densities 
and more numerous constraints 
on the coverage fractions, ionizations and abundances (\S3). 
BEL studies should include more of the 
weaker lines, such as OVI~\lam 1034, CIII~\lam 977, 
NIII~\lam 991 and the intercombination lines, whenever 
possible (\S2). Theoretical analysis of the 
FeII/MgII emission ratios, in particular, is needed to test the 
tentative conclusion for high Fe/Mg abundances. This and other 
BEL results should be tested further by examining the same lines 
(or same elements) in intrinsic NAL systems. The steady 
improvement in our observational capabilities at all wavelengths 
will provide many more diagnostic opportunities.

Below are some specific issues that new studies might address. 

1) More data at high redshifts will  
constrain better the epoch and extent of early star formation 
associated with QSOs. 

2) Reliable measurements of Fe/$\alpha$ will further constrain 
the epoch of first star formation and, perhaps, the cosmology 
via the $\sim$1~Gyr enrichment clock. 

3) Better estimates of the metal-to-metal ratios generally 
will reveal more specifics of the star formation histories 
via comparisons to well-studied galactic environments 
and theoretical nucleosynthetic yields.

4) Abundances for QSOs spanning a wide range of luminosities and 
redshifts will isolate any evolutionary (redshift) trends and test 
the tentative luminosity--$Z$ relationship. This relationship might 
prove to be a useful indicator of the total masses or 
densities of the local stellar populations by analogy with the 
mass--$Z$ trend in nearby galaxies. 

5) The range of QSO metallicities at a given redshift and luminosity 
will help constrain the extent of star formation 
occurring before QSOs 
turn on or become observable. Are there any low metallicity QSOs? 

6) Combining the QSO abundances with direct imaging studies 
of their host galaxies should test ideas about the chemical enrichment 
and help us interpret data at the highest redshifts 
where direct imaging is (so far) not possible. For example, are QSOs 
in large galaxies (e.g. giant ellipticals) more metal-rich than others? 

7) Correlations between the abundances and other properties of 
QSOs, such 
as radio-loudness or UV--X-ray continuum shape, might reveal new 
environmental factors in the enrichment or systematic 
uncertainties in our abundance derivations. 

8) Observations with wide wavelength coverage would allow us to 
compare abundances derived from the narrow emission lines (in the 
rest-frame optical) to BEL and NAL data 
in the same objects. These diverse diagnostics 
might provide crude abundance maps of QSO host galaxies. 

9) How do QSO abundances compare to their low-redshift 
counterparts, the Seyfert galaxies and active galactic 
nuclei (AGNs)? Low-redshift metallicities might be 
less than the QSOs due to recent mergers or gaseous infall. 

\bigskip\bigskip\bigskip

\noindent ACKNOWLEDGEMENTS 

\noindent We are grateful to G Burbidge for his patience and encouragement. 
We also thank KT Korista and JC Shields for helpful comments on this 
manuscript, and TA Barlow, N Arav and VT Junkkarinen for useful 
discussions. GF thanks the Canadian Institute for Theoretical Astrophysics 
for their hospitality during a sabbatical year, and acknowledges support 
from the Natural Science and Engineering Research Council of Canada 
through CITA. The work of FH was supported by NASA grant NAG 5-3234. 
Research in nebular astrophysics at the University of Kentucky is supported
by the NSF through grant 96-17083 and by NASA through its ATP (award NAG
5-4235) and LTSA programs.

\newpage
\section{Literature Cited}
\parskip=0pt
\leftskip=0.5in
\parindent=-0.5in

\def  \ARAA   {{\it Annu. Rev. Astron. Astrophys.} }
\def  \ApJ    {{\it Ap. J.} }
\def  \ApJS   {{\it Ap. J. Suppl.} }
\def  \AJ     {{\it Astron. J.} }
\def  \AA     {{\it Astron. Astrophys.} }
\def  \MNRAS  {{\it MNRAS} }
\def  \PASP   {{\it PASP} }

Aldcroft TL, bechtold J, Elvis M. 1994, \ApJS 93, 1

Anderson SF, Weymann RJ, Foltz CB, Chaffee FH.
 1987, \AJ 94, 278

Arav N. 1996, \ApJ 465, 617

Arav N. 1997, in Mass Ejection From AGN, eds. R Weymann, I Shlosman, 
N Arav, ASP Conf. Series, 128, 208

Arav N, Barlow TA, Laor A, Sargent WLW, Blandford RD. 1998, \MNRAS 297, 990

Arav N, Korista KT, de Kool M, Junkkarinen VT, Begelman MC. 1999, 
\ApJ in press (A99)

Aretxaga I, Terlevich RJ, Boyle BJ. 1998, \MNRAS 296, 643

Arimoto N  Yoshii Y. 1987, \AA 173, 23

Artymowicz P, Lin DNC, Wampler EJ. 1993, \ApJ 409, 592

Bahcall JN, Kirhakos S, Saxe DH, Schneider DP. 1997, \ApJ 
479, 642

Bahcall JN, Sargent WLW, Schmidt M. 1967, \ApJ 149, L11

Baldwin JA. 1977a, \MNRAS 178, 67P

Baldwin JA. 1977b, \ApJ 214, 679

Baldwin JA, \& Netzer H. 1978, \ApJ 226, 1

Baldwin JA, Ferland GJ, Korista KT, Carswell RF, 
Hamann F, \etal 1996, \ApJ 461, 664

Baldwin JA, Ferland GJ, Korista KT, Verner D. 1995, \ApJ 455, L119

Barbuy B. 1988, \AA 191, 121

Barger AA, Aragon-Salamanca A, Smail I, Ellis RS, Couch WJ, 
\etal 1998a, \ApJ 501, 522

Barger AA, Cowie LL, Trentham N, Fulton E, Hu EM, \etal 1998b, 
preprint (astro-ph/9809299)

Barlow TA. 1993, Ph.D. Dissertation, University of California 
-- San Diego

Barlow TA. 1998, private comm.

Barlow TA, Hamann F, Sargent WLW. 1997, in 
Mass Ejection From AGN, eds. R Weymann, I Shlosman, N Arav, 
ASP Conf. Series, 128, 13

Barlow TA, Junkkarinen VT. 1994, BAAS, 26, 1339

Barlow TA, Junkkarinen VT, Burbidge EM, Weymann RJ, 
Morris SL, \etal 1992, \ApJ 397, 81

Barlow TA, Sargent WLW. 1997, \AJ 113, 136

Barthel PD, Tytler DR, Vestergaard M. 1997,  in 
Mass Ejection From AGN, eds. R Weymann, I Shlosman, N Arav, 
ASP Conf. Series, 128, 48

Baugh CM, Cole S, Frenk CS, Lacey CG. 1998, \ApJ 498, 504

Bender R, Burstein D, Faber SM. 1993, \ApJ 411, 153

Bergeron J, Boiss\'e P. 1986, \AA 168, 6

Bergeron J, Kunth D. 1983, \MNRAS 205, 1053

Bergeron J, Petitjean P, Sargent WLW, Bahcall JN, Boksenberg A, 
\etal 1994, \ApJ 436, 33

Bergeron J  Stasi\'nska G. 1986, \AA 169, 1

Bernardi M, Renzini A, da Costa LN, Wegner G, Victoria M, 
\etal 1998, preprint (astro-ph/9810066)

Bica E, Alloin D, Schmidt AA. 1990, \AA 228, 23

Bica E, Arimoto N,   Alloin D. 1988, \AA 202, 8

Boyce PJ, Disney MJ, Bleaken DG. 1999, \MNRAS 302, 39

Boyle B. 1990, \MNRAS 243, 231

Boyle BJ, Terlevich RJ. 1998, \MNRAS 293, 49

Branch D. 1998, \ARAA 36, 17

Bressan A, Chiosi C, Tantalo R. 1996, \AA 311, 425

Brotherton MS, Van Breugel W, Smith RJ, Boyle BJ, Shanks T. 1998b, 
\ApJ 505, 7

Brotherton MS, Wills BJ, Dey A, Van Breugal W, Antonucci R. 1998a, 
\ApJ 501, 110

Brunner H, Mueller C, Friedrich P, Doerrer T, Staubert R, \etal 
1997, \AA 326, 885

Bruzual G, Barbuy B, Ortolani S, Bica E, Cuisinier F. 1997, 
\AJ 114, 1531

Bruzual G, Magris G. 1997, in the STScI Symp on the Hubble Deep Field, 
(astro-ph/9707154)

Burbidge G, Burbidge M. 1967, in Quasi-Stellar Objects, 
WH Freeman and Co.

Burles S,  Tytler D. 1996, \ApJ 460, 584

Carroll SM, Press WH. 1992, \ARAA 30, 499

Carballo R, Sanchez SF, Gonzalas-Serrano JI, Benn CR, Vigotti M. 
1998, \AJ 115, 1234

Castro S, Rich RM, McWilliam A, Ho LC, Spinrad H, \etal 1996, 
\AJ 111, 2439

Cavaliere A, Vittorini V. 1998, in The Young Universe, eds. S D'Odorico, 
A Fontana, E Giallongo, ASP Conf. Ser., 146, 28 

Chatzichristou ET, Vanderriest C, Jaffe W. 1999, \AA 343, 407

Cohen MH, Ogle PM, Tran HD, 
Vermeulen RC, Miller JS, \etal 1995, \ApJ 448, L77

Collin S. 1998, preprint

Connolly AA, Szalay AS, Dickenson M, SubbaRao MU, Brunner RJ. 1997, 
\ApJ 486, L11

Cozial R, Contini T, Davoust E, Consid\`ere S. 1997, \ApJ 481, L67

Davidson K. 1977, \ApJ 218, 20

Davidson K, Netzer H. 1979, Rev. Mod. Phys., 51, 715, (DN79).

de Freitas Pacheco JA. 1996, \MNRAS 278, 841

Dey A, Spinrad H, Stern D, Graham JR, Chaffee FH.
 1998, \ApJ 498, L93

Djorgovski SG. 1998, in Fundamental Parameters of Cosmology, 
ed. Y Giroud-H\`eraud, (Gif sur Yvette:Editions Fronti\`eres), 
in press

Edmunds MG. 1992, in Elements and the Cosmos, eds. MG Edmunds, R Terlevich, 
(Cambridge Univ. Press:New York), p.289

Ellis R, Smail I, Dressler A, Couch WJ, Oemler WJ, \etal 
1997, \ApJ 483, 582

Faber SM. 1973, \ApJ 179, 423

Faber SM, Wegner G, Burstein D, Davies RL, Dressler A, 
\etal 1989, \ApJS 69, 763

Feltzing S, Gilmore G. 1998, in Galaxy Evolution: Connecting the 
Distant Universe with the Local Fossil Record, p. 71

Ferland GJ. 1999, in Quasars as Standard Candles for Cosmology, 
eds. J Baldwin, GJ Ferland, KT Korista, in press

Ferland GJ, Baldwin JA, Korista KT, Hamann F, Carswell 
RF, \etal 1996, \ApJ 461, 683

Ferland GJ, Korista KT, Verner DA, Ferguson JW, Kingdon JB, \etal 
1998, \PASP 110, 761

Ferland GJ, Persson SE. 1989, \ApJ 347, 656

Ferland GJ, Peterson BM, Horne K, Welsch WF,   
Nahar SN. 1992, \ApJ 387, 95

Ferland GJ, Shields GA. 1985, in Astrophysics of Active Galaxies 
and Quasi-Stellar Objects, ed. JS Miller, Univ. Sci. Books, p. 157

Fisher D, Franx M, Illingworth G. 1995, \ApJ 448, 119

Foltz CB, Chaffee F, Weymann RJ,   Anderson SF.
1988, in QSO Absorption Lines: Probing the 
Universe,  eds. JC Blades, DA Turnshek, CA Norman (Cambridge: 
Cambridge Univ Press), p. 53

Foltz CB, Weymann RJ, Peterson BM, Sun L, Malkan MA, \etal 
1986, \ApJ 307, 504

Franceschini A, Gratton R. 1997, \MNRAS 286, 235

Francis PJ, Koratkar A. 1995, \MNRAS 274, 504

Franx M, Illingworth GD. 1990, \ApJ 359, L41

Franx M, Illingworth GD, Kelson DD, Van Dokkum PG, 
  Tran K. 1997, \ApJ 486, 75

Friaca ACS, Terlevich RJ. 1998, \MNRAS 298, 399

Friaca ACS, Terlevich RJ. 1999, \MNRAS in press

Frisch H. 1984, in Methods in Radiative Transfer, ed. W Kalkofen 
(Cambridge:Cambridge Univ. Press), 65

Gallagher SC, Brandt WN, Sambruna RM, Mathur S, Yamasaki N. 1999, 
\ApJ in press

Ganguly R, Eracleous M, Charlton JC, Churchill CW. 1999, \AJ 
in press

Garnett DR. 1990, \ApJ 363, 142

Gaskell CM, Shields GA, and Wampler EJ. 1981, \ApJ 249, 443

Geisler D, Friel DE. 1992, \AJ 104, 128

Giavalisco M, Steidel CC,   Macchetto FD. 1996, \ApJ 470, 189

Gnedin NY, Ostriker JP. 1997, \ApJ 486, 581

Goodrich RW, Miller JS. 1995, \ApJ 448, L73

Gorgas J, Efstathiou G, Arag\'on Salamanca AA. 1990, \MNRAS 245, 217

Grandi SA. 1981, \ApJ 251, 451

Gratton RG. 1991, in Evolution of Stars: The Photospheric 
Abundance Connection, eds. G. Michaud, AV Tutukov, 
IAU Symp. 145, (Montreal:Univ. Montreal), 27

Green PJ, Aldcroft TL, Mathur S, Shartel N. 1997, \ApJ 484, 135

Green PJ, Mathur S. 1996, \ApJ 462, 637

Greggio L, and Renzini A. 1983 \AA 118, 217

Grevesse N,   Anders E. 1989, in Cosmic Abundances of Matter, 
AIP Conf. Proc. 183, ed. CI Waddington (New York:AIP), 1

Grillmair CJ, Turnshek DA. 1987, in QSO Absorption Lines: Probing the 
Universe, Poster Papers, eds. JC Blades, C Norman, DA Turnshek, 
p. 1

Guilloteau S, Omont A, McMahon RG, Cox P, Petitjean P. 1997, 
\AA 328, L1

Haas M, Chini R, Maisenheimer K, Stickel M, Lemke D, \etal 1998, 
\ApJ 503, L109

Haehnelt MG, Natarajan P,  Rees MJ.  1998, \MNRAS 300, 817

Haehnelt MG, Rees MJ.  1993, \MNRAS 263, 168

Haiman Z, Loeb A. 1998, \ApJ 503, 505

Hamann F. 1997, \ApJS 109, 279 (H97)

Hamann F. 1998, \ApJ 500, 798

Hamann F, Barlow TA, Beaver EA, Burbidge EM, Cohen 
RD, \etal 1995, \ApJ 443, 606

Hamann F, Barlow TA, Junkkarinen V. 1997c, \ApJ 478, 87 

Hamann F, Barlow TA, Cohen RD, Junkkarinen V, Burbidge EM. 
1997b, \ApJ 478, 80 (H97b)

Hamann F, Barlow TA, Cohen RD, Junkkarinen V, Burbidge EM. 
1997e, in Mass Ejection From AGN, eds. R Weymann, I Shlosman, 
N Arav, ASP Conf. Series, 128, 187

Hamann F, Beaver EA, Cohen RD, Junkkarinen V, Lyons RW, \etal. 1997d,
\ApJ 488, 155

Hamann F, Chaffee R, Weymann RJ, Barlow TA, Junkkarinen VT. 1999b, 
in prep.

Hamann F, Cohen RD, Shields JC, Burbidge EM, Junkkarinen VT, \etal 
1998, ApJ, 496, 761

Hamann F,   Ferland GJ. 1992, ApJL, 391, L53

Hamann F,   Ferland GJ. 1993a, Rev. Mex. Astr. Astrof., 26, 53

Hamann F,   Ferland GJ. 1993b, \ApJ 418, 11, (HF93b)

Hamann F,   Korista KT. 1996, \ApJ 464, 158 

Hamann F, Korista KT, Ferland GJ. 1999a, in prep.

Hamann F, Korista KT, Morris SL. 1993, \ApJ 415, 541

Hamann F, Shields JC, Cohen RD, Junkkarinen VT, Burbidge EM. 
1997a, in Emission Lines From Active Galaxies: New Methods and Techniques, 
IAU Col. 159, eds. BM Peterson, F-Z Cheng, AS Wilson, ASP Conf. Ser., 
113, 96

Hartquist TW,  Snijders MAJ. 1982, Nature, 299, 783

Heap SR, Brown TM, Hubeny I, Landsman W, Yi S, \etal 1998, \ApJ 492, L131

Hines DC, Low FJ, Thompson RI, Weymann RJ, Storrie-Lombardi LJ. 1999, 
\ApJ 512, 140

Hines DC, Wills BJ. 1995, \ApJ 448, L69

Hu EM, Cowie LL, McMahon RG. 1998, \ApJ 502, L99

Idiart TP, De Freitas Pacheco JA,   Costa RDD.
 1996, \AJ 112, 2541

Isaac KG, McMahon RG, Hills RE, Withington S. 1994, \MNRAS 269, 28

Isotov YI, Thuan TX. 1999, \ApJ in press

Israelian G, L\'opez RJG, Rebolo R. 1998, \ApJ 507, 805

Ivison RJ, Dunlop JS, Hughes DH, Archibald EN, Stevens JA, \etal 
1998, \ApJ 494, 211

Jablonka P, Alloin D,   Bica E. 1992, \AA 260, 97

Jablonka P, Martin P,   Arimoto N. 1996, \AJ 112, 1415

Johansson S, Jordan C. 1984, \MNRAS 210, 239

Jenkins EB. 1996, \ApJ 471, 292

Jin L, Arnett WD, Chakrabarti SK. 1989, \ApJ 336, 572

Junkkarinen VT. 1980, Ph.D. Dissertation, University of California 
-- San Diego

Junkkarinen VT. 1998, private comm.

Junkkarinen VT, Beaver EA, Burbidge EM, Cohen RC, 
Hamann F, \etal 1997, in 
Mass Ejection From AGN, eds. R Weymann, I Shlosman, N Arav, 
ASP Conf. Series, 128, 220

Junkkarinen VT, Burbidge EM, Smith HE. 1983, \ApJ 265, 
51

Junkkarinen VT, Burbidge EM, Smith HE. 1987, \ApJ 317, 
460

Katz N, Quinn T, Bertschinger E, Gelb JM. 1994, \MNRAS 270, L71

Kauffmann G. 1996, \MNRAS 281, 487

King JR. 1993, \AJ 106, 1206

Kinney AL, Rivolo AR, Koratkar AR. 1990, \ApJ 357, 338

Kirkman D, Tytler D. 1997, \ApJ 489, L123

Kobulnicky HA, Skillman ED. 1998, \ApJ 497, 601

Kodama T, Arimoto N. 1997, \AA 320, 41

K\" oppen J,   Arimoto N. 1990, \AA 240, 22

Korista KT, Alloin D, Barr P, Clavel J, Cohen RD, \etal 1995, 
\ApJS 97, 285

Korista KT, Baldwin JA, Ferland GJ. 1998, \ApJ 507, 24

Korista KT, Baldwin JA, Ferland GJ, Verner D. 1997b, \ApJS 108, 401

Korista KT, Ferland GJ, Baldwin BA. 1997a, \ApJ 487, 555

Korista KT, Hamann F, Ferguson J, Ferland GJ. 1996, 
\ApJ 461, 641

Korista KT, Voit GM, Morris SL, Weymann RJ. 1993, \ApJS 88, 357

Korista KT, Weymann RJ, Morris SL, Kopko M, Turnshek
DA, \etal 1992, \ApJ 401, 529

Kormendy J, Bender R, Evans AS, Richstone D. 1998, 
\AJ 115, 1823

Krolik J, Voit GM. 1998, \ApJ 497, L5

Kundt W. 1996, {\it Astrophys. Sp. Sci.} 235, 319

Kundtschner H, Davies RL. 1997, \MNRAS 295, L29

Kwan J. 1990, \ApJ 353, 123

Kwan J, Krolik J. 1981, \ApJ 250, 478, (KK81)

Laor A. 1998, in Quasars as Standard Candles for Cosmology, eds. 
J Baldwin, GJ Ferland, KT Korista, in press

Laor A, Bahcall JN, Jannuzi BT, Schneider DP, Green RF. 1995, \ApJS 99, 1

Laor A, Jannuzi BT, Green RF, Boroson TA. 1997, \ApJ 489, 656

Larson RJ. 1974, \MNRAS 169, 229

Loeb A. 1993, \ApJ 404, L37

Loeb A, Rasio FA. 1994, \ApJ 432, 52

Lowenthal JD, Koo DC, Guzman R, Gallego J, Phillips AC. 1997, \ApJ 481, 673

Lu L, Sargent WLW, Barlow TA. 1998, \AJ 115, 55

Lu L, Sargent WLW, Barlow TA, Churchhill CW, Vogt SS. 1996, \ApJS 107, 475

Lu L, Savage BD. 1993, \ApJ 403, 127

Madau P, Ferguson HC, Dickenson ME, Giavalisco M, Steidel CC, 
\etal 1996, \MNRAS 283, 1388

Magain P. 1989, \AA 209, 211

Magorrian J, Tremaine S, Richstone D, Bender R, Bower G, 
\etal 1998, \AJ 115, 2285

Mathews WG, Capriotti ER. 1985, in Astrophysics of Active Galaxies 
and Quasi-Stellar Objects, ed. JS Miller, Univ. Sci. Books, p. 183

Matteucci F. 1994, \AA 288, 57

Matteucci F, Brocato E. 1990, \ApJ 365, 539

Matteucci F, Francois P. 1989, \MNRAS 239, 885

Matteucci F, Greggio L. 1986, \AA 154, 279

Matteucci F, Tornamb\`e A. 1987. \AA 185, 51

McCarthy PJ. 1993, \ARAA 31, 639

McLeod KK, Rieke GH 1995, \ApJ 454, L77

McLeod KK, Rieke GH, Storrie-Lombardi LJ. 1999, \ApJ 511, L67

McLure RJ, Dunlop JS, Kukula MJ, Baum SA, 
O'Dea CP, \etal 1998, preprint (astro-ph/9809030)

McWilliam A,   Rich RM. 1994, \ApJS 91, 749

Mihalas D. 1978, Stellar Atmospheres, WH Freeman \& Co., 
(San Francisco:Univ. Chicago Press)

Miller J, Tran H,   Sheinis A. 1996, BAAS, 28, 1031

Minniti D, Olszewski EW, Liebert J, White SD, Hill JM, 
\etal 1995, \MNRAS 277, 1293

Miralda-Escud\'e J, Rees MJ. 1997, \ApJ 478, L57

M\"oller P, Jakobsen P, Perryman MAC. 1994, \AA 287, 719

Morris SL, Weymann RJ, Foltz 
CB, Turnshek DA, Shectman S, \etal 1986, \ApJ 310, 40

Murray N, Chiang J, Grossman SA, Voit GM. 1995, \ApJ 451, 498

Mushotzky RF, Loewenstein M.  1997, \ApJ 481, L63

Netzer H. 1990, in Active Galactic Nuclei, eds. RD Blandford, 
H Netzer, L Woltjer, (Berlin:Springer), 57

Netzer H, Brotherton MS, Wills BJ, Han M, Baldwin JA, \etal 
1995, 448, 27

Netzer H, Wills BJ. 1983, \ApJ 275, 445

Nissen PE, Gustafsson B, Edvardsson B, Gilmore G. 1994, \AA 285, 440

Omont A, Petitjean P, Guilloteau S, McMahon RG, Solomon PM. 
1996, Nature, 382, 428

Ortolani S, Renzini A, Gilmozzi R, Marconi G, Barbuy B, \etal 
1996, in Formation of the Galactic Halo...Inside and Out, 
Eds. H Morrison, A Sarajedini, ASP Conf. Ser., 92, 96

Osmer PS. 1980, \ApJ 237, 666

Osmer PS. 1998, in The Young Universe, eds. S D'Odorico, A Fontana, 
E Giallongo, ASP Conf. Ser., 146, 1

Osmer PS, Porter AC, Green RF. 1994, \ApJ 436, 678

Osmer PS, Shields, JC. 1999, in Quasars as Standard Candles for Cosmology, 
eds. J Baldwin, GJ Ferland, KT Korista, in press

Osmer PS, Smith MG. 1976, \ApJ 210, 276

Osmer PS, Smith MG. 1977, \ApJ 213, 607

Osterbrock DE. 1977, \ApJ 215, 733

Osterbrock DE. 1989, Astrophysics of Gaseous Nebulae and 
Active Galactic Nuclei, University Science Press 

Peimbert M. 1967, \ApJ 150, 825

Penston M. 1987, \MNRAS 229, 1P

Peterson BM. 1993, \PASP 105, 1084

Petitjean P, Rauch M, Carswell RF. 1994, \AA 291, 29

Petitjean P, Srianand R. 1999, \AA in press

Pettini M,  King DL, Smith LJ, Hunstead RW. 1997, \ApJ 
486, 665

Phillips MM. 1977, \ApJ 215, 746

Phillips MM. 1978, \ApJ 226, 736

Prochaska JX, Wolfe AM. 1998, \ApJ 507, 113

Prochaska JX, Wolfe AM. 1999, \ApJ in press

Rauch M. 1998, \ARAA 36, 267

Rauch M, Haehnelt MG, Steinmetz M. 1997, \ApJ 481, 601

Renzini A. 1997, \ApJ 488, 35

Renzini A. 1998, in The Young Universe, eds. S D'Odorico, A Fontana, 
E Giallongo, ASP Conf. Ser., 146, 298

Rich RM. 1988, \AJ 95, 828

Rich RM. 1990, \ApJ 362, 604

Richards GT, York DG, Yanny B, Kollgaard RI, Laurent-Muehleisen SA, 
\etal 1999, \ApJ in press

Saikia D, Kulkarni AR. 1998, \MNRAS 298, L45

Sansom AE, Proctor RN. 1998, \MNRAS 297, 953

Savage BD, Sembach KR. 1991, \ApJ 379, 245

Savage BD, Tripp TM, Lu L. 1998, \AJ 115, 436

Savaglio S, Cristiani S, D'Odorico S, Fontana A, Giallong E, \etal 
1997, \AA 318, 347

Savaglio S, D'Odorico S,  M\"o ller P. 1994, \AA 281, 331

Scalo JM.  1990, in Windows on Galaxies,  eds. G Fabbiano, 
JS Gallagher, A Renzini, (Dordrecht:Kluwer Academic Publishers) 
p. 125

Schmidt GD, Hines DC. 1999, \ApJ in press

Schneider DP. 1998, in Science with NGST, eds. EP Smith, A Koratkar, 
ASP Conf. Ser., 133, 106

Schneider DP, Schmidt M, and Gunn JE. 1991, \AJ 102, 837

Searle L, Zinn R. 1978, \ApJ 225, 357

Shaver PA, Hook IM, Jackson CA, Wall JV, Kellerman KI. 1998, 
in Highly redshifted Radio Lines, eds. C Carilli, S Radford, 
K Menton, G. Langston, (PASP:San Francisco), in press

Shields GA.  1976, \ApJ 204, 330

Shields GA. 1996, \ApJ 461, L9

Shields JC, Ferland GJ, Peterson, BM. 1995, \ApJ 441, 507

Shields JC, Hamann F, Foltz CB, Chaffee FH. 1997, in Emission 
Lines in Active Galaxies, eds. BM Peterson, F-Z Cheng, 
ASP Conf. Ser., 113, 118

Shklovskii IS. 1965, Sov. Astr., 8, 635

Sigut TAA, Pradhan AK. 1998, \ApJ 499, L139

Sil'chenko OK, Burenkov AN, Vlasyuk VV. 1998, \AA 337, 349

Silk J, Rees MJ. 1998, \AA 331, L1

Songalia A, Cowie LL. 1996, \AJ 112, 335

Spaans M, Corollo CM. 1997, \ApJ 482, L93

Spinrad H, Dey A, Stern D, Dunlop J, Peacock J, 
\etal 1997, \ApJ 484, 581

Srianand R, Shankaranarayanan S. 1999, \ApJ in press 

Stanford SA, Eisenhardt PR, Dickenson M. 1998, 
\ApJ 492, 461

Steidel CC, Adelberger KL, Dickenson M, Giavalisco M, 
Pettini M, \etal 1998, \ApJ 492, 428

Steidel CC, Adelberger KL, Giavalisco M, Dickenson M,  
Pettini M. 1999, \ApJ in press

Taniguchi Y, Arimoto N, Murayama T, Evans AS, Sanders DB, 
\etal 1997, in Quasar Hosts, eds. DL Clements, I Perez-Fouron, 
ESO-IAC Conf. Pro., p. 127

Taniguchi Y, Ikeuchi S, Shioya Y. 1999, \ApJ 514, L12

Tantalo R, Chiosi C, Bressan A. 1998, \AA 333, 419

Telfer RC, Kriss GA, Zheng W, Davidsen AF, Green RF. 1999, 
\ApJ in press

Terndrup DM, Sadler EM, Rich RR. 1995, \AJ 110, 1774

Terlevich RJ, Boyle BJ. 1993, \MNRAS 262, 491

Thomas D, Greggio L, \& Bender R. 1998, \MNRAS in press

Thompson KL, Hill GJ, Elston R. 1999, \ApJ in press

Thuan TX, Izotov YI, Lipovetsky VA. 1995, \ApJ 445, 108

Thurston TR, Edmunds MG, Henry RB. 1996, \MNRAS 283, 990

Tiede GP, Frogel JA, Terndrup DM. 1995, \AJ 110, 2788

Tinsley B. 1979, \ApJ 229, 1046

Tinsley B. 1980, Fund. of Cosmic Phys., 5, 287

Trager SC, Faber SM, Dressler A, Oemler A. 1997, \ApJ 485, 92

Tripp TM, Lu L, Savage BD. 1996, \ApJS 102, 239

Tripp TM, Lu L, Savage BD. 1997, \ApJS 112, 1

T\"urler M, Courvoisier TJ-L. 1997, \AA 329, 863

Turner EL.  1991, \AJ 101, 5

Turnshek DA. 1981, Ph.D. Dissertation, University of Arizona

Turnshek DA. 1984, 280, 51

Turnshek DA. 1986, in Quasars, eds. G. Swarup, VK Kapahi, IAU Symp. 
No. 119, (Dordrecht: Reidel), p. 317

Turnshek DA, Briggs FH, Foltz CB, Grillmair CJ, Weymann RJ. 
1987, in QSO Absorption Lines: Probing the 
Universe, Poster Papers, eds. JC Blades, C Norman, DA Turnshek, 
p. 1

Turnshek DA. 1988, in QSO Absorption Lines: Probing the 
Universe,  eds. JC Blades, C Norman, DA Turnshek, (Cambridge: 
Cambridge Univ Press), p. 17

Turnshek DA. 1994, in QSO Asborption Lines, ed. G Meylan, 
(Springer-Verlag), p. 223

Turnshek DA. 1997, in Mass Ejection From AGN,  eds. R Weymann, 
I Shlosman, N Arav, ASP Conf. Series, 128, 193

Turnshek DA, Kopko M, Monier E, Noll D, Espey B, \etal 
1996, \ApJ 463, 110

Turnshek DA, Monier EM, Christopher JS, Espey BR. 1997, 
\ApJ 476, 40

Tytler D, Fan XM, Burles S, Cottrell L, Davis C, \etal 1995, 
in QSO Absorption Lines, ed. G Meylan, (Garching:ESO), 289

Uomoto A. 1984, ApJ 284, 497

Van Dokkum PG, Franx M, Kelson DD,   Illingworth GD.
 1998, \ApJ 504, L17

Van Zee L, Skillman ED, Salzer JJ. 1998, \ApJ 497, L1

Vazdekis A, Peletier RF, Beckman JE,   Casuso 
E. 1997, \ApJS 111, 203

Verner EM, Verner DA, Korista KT, Ferguson JW, Hamann F, \etal 
1999, \ApJ in press

Vila-Costas MB, Edmunds MG. 1993, \MNRAS 265, 199

Voit GM, 1997, in Mass Ejection From AGN,  eds. R Weymann, 
I Shlosman, N Arav, ASP Conf. Series, 128, 200

Voit GM, Weymann RJ, Korista KT. 1993, ApJ 413, 95

Wampler EJ, Bergeron J, Petitjean P. 1993, \AA 273, 15

Wampler EJ, Chugai NN, Petitjean P. 1995, \ApJ 443, 586

Warren SJ, Hewett PC, Osmer PS. 1994, \ApJ 421, 412

Weymann RJ. 1994, in QSO Asborption Lines, ed. G Meylan, 
(Springer-Verlag), p. 213

Weymann RJ. 1997, in Mass Ejection From AGN,  eds. R Weymann, 
I Shlosman, N Arav, ASP Conf. Series, 128, 3

Weymann RJ, Foltz C. 1983, in Quasars and Gravitational Lenses, 
24th Liege Int. Astrophy. Coll., (Univ. de Liege:Belgium), p. 538

Weymann RJ, Turnshek DA, Christiansen WA. 1985, 
in Astrophysics of Active Galaxies and Quasi-Stellar Objects, 
ed. J Miller, (Mill Valley, CA: University Science Books), 185

Weymann RJ, Morris SL, Foltz CB, Hewett PC. 1991,
\ApJ 373, 23 

Weymann RJ, Carswell RF, Smith MGA. 1981, \ARAA 19, 41

Weymann RJ, Stern D, Bunker A, Spinrad H, Chaffee FH, \etal 
1998, \ApJ 505, L95

Weymann RJ, Williams RE, Peterson BM, Turnshek 
DA. 1979, \ApJ 218, 6 19

Wheeler JC, Sneden C, Truran JW. 1989, \ARAA
27, 279 

Williams RE. 1971, \ApJ 167, L27

Williams RE, Strittmatter PA, 
Carswell RF,  Craine ER. 1975, \ApJ 202, 296

Williams RE,  Weymann RJ. 1976, \ApJ 207, L143

Will BJ, Netzer H. 1983, \ApJ 275, 445

Wills BJ, Netzer H, and Wills D. 1985, \ApJ 288, 94

Wills BJ, Thompson KL, Han M, Netzer H, Wills D, \etal 1995, \ApJ 447, 139

Worthey G, Faber SM, Gonzalez J. 1992, \ApJ 398, 69

Yoshii Y, Tsujimoto T, Kawara K. 1998, \ApJ 507, L113

Yoshii Y, Tsujimoto T, Nomoto K. 1996, \ApJ 462, 266

Young P, Sargent WLW, and Boksenberg A. 1982, \ApJS 48, 455

Zaritsky D, Kennicutt RC, Huchra JP. 1994, \ApJ 420, 87

Zheng W, Kriss GA, Davidsen AF. 1995, \ApJ 440, 606

Zheng W, Kriss GA, Telfer RC, Grimes JP, Davidsen AF. 1997, \ApJ 475, 469

Ziegler BL, Bender R. 1997, \MNRAS 291, 527

\end{document}